\documentclass[a4paper,twocolumn,accepted=2021-03-29]{quantumarticle}
\pdfoutput=1
\usepackage[T1]{fontenc}

\usepackage[utf8]{inputenc}
\usepackage{amsmath,amssymb}
\usepackage{amsfonts}
\usepackage{epsfig,pstricks,graphicx}
\usepackage{bm}
\usepackage{color}
\def\id{{\rm 1\kern-.22em l}}
\usepackage{enumerate}

\usepackage[unicode]{hyperref}
\usepackage{array}
\usepackage{booktabs}

\newcolumntype{C}[1]{>{\centering\let\newline\\\arraybackslash\hspace{0pt}}m{#1}}

\usepackage{float}

\usepackage[numbers,sort&compress]{natbib}

\begin{document}

\title{Experimentally friendly approach towards nonlocal correlations in multisetting $N$-partite Bell scenarios} 

\author{Artur Barasi\'nski}
\email{artur.barasinski@uwr.edu.pl}
\affiliation{Institute of Theoretical Physics, Uniwersity of Wroclaw, Plac Maxa Borna 9, 50-204 Wrocław, Poland}
\affiliation{RCPTM, Joint Laboratory of Optics of Palack\'{y} University and Institute of Physics of CAS, Faculty of Science, Palack\'{y} University, 17. Listopadu 12, 771 46 Olomouc, Czech Republic}

\author{Anton\'{i}n \v{C}ernoch}
\affiliation{Institute of Physics of the Czech Academy of Sciences, Joint Laboratory of Optics of Palacký University and Institute of Physics of CAS, 17. listopadu 50A, 772 07 Olomouc, Czech Republic}

\author{Wiesław Laskowski}
\affiliation{Institute of Theoretical Physics and Astrophysics, Faculty of Mathematics, Physics and Informatics,
University of Gdańsk, 80-308 Gdańsk, Poland}
\affiliation{International Centre for Theory of Quantum Technologies, University of Gdańsk, 80-308 Gdańsk, Poland}

\author{Karel Lemr}
\email{k.lemr@upol.cz}
\affiliation{RCPTM, Joint Laboratory of Optics of Palack\'{y} University and Institute of Physics of CAS, Faculty of Science, Palack\'{y} University, 17. Listopadu 12, 771 46 Olomouc, Czech Republic}

\author{Tamás Vértesi}
\affiliation{MTA Atomki Lendület Quantum Correlations Research Group, Institute for Nuclear Research, P.O. Box 51, H-4001 Debrecen, Hungary}

\author{Jan Soubusta}
\affiliation{RCPTM, Joint Laboratory of Optics of Palack\'{y} University and Institute of Physics of CAS, Faculty of Science, Palack\'{y} University, 17. Listopadu 12, 771 46 Olomouc, Czech Republic}

\begin{abstract}
In this work, we study a recently proposed operational measure of nonlocality by Fonseca and Parisio~[Phys. Rev. A 92 , 030101(R) (2015)] which describes the probability of violation of local realism under randomly sampled observables, and the strength of such violation as described by resistance to white noise admixture. While our knowledge concerning these quantities is well established from a theoretical point of view, the experimental counterpart is a considerably harder task and very little has been done in this field. It is caused by the lack of complete knowledge about the facets of the local polytope required for the analysis. In this paper, we propose a simple procedure towards experimentally determining both quantities for $N$-qubit pure states, based on the incomplete set of tight Bell inequalities. We show that the imprecision arising from this approach is of similar magnitude as the potential measurement errors. We also show that even with both a randomly chosen $N$-qubit pure state and randomly chosen measurement bases, a violation of local realism can be detected experimentally almost $100\%$ of the time. Among other applications, our work provides a feasible alternative for the witnessing of genuine multipartite entanglement without aligned reference frames.
\end{abstract}

\maketitle

\section{Introduction}

Nonlocality is arguably one of the most striking aspects of quantum mechanics, dramatically defying our intuition about time and space \cite{Bellphysics1_1964}. Although this feature was initially thought to be an evidence of the incompleteness of the quantum theory \cite{Einsteinpr47_1935}, there is today overwhelming experimental evidence that nature is indeed nonlocal \cite{Aspectnature398_1999}. 
Nonlocality also plays a central role in quantum information science and has been recognized an essential resource for quantum information tasks \cite{Brunnerrmp86_2014}, for instance, quantum key distribution \cite{Bennet_proc, Ekertprl67_1991}, communication complexity \cite{Buhrmanrmp82_2010}, randomness generation \cite{Pironionature464_2010}, and device-independent information processing  \cite{Acinprl98_2007,Rabeloprl107_2011}. All such device-independent applications require states that strongly violate Bell inequalities. However, the concept of “strength of violation” is controversial in the literature \cite{Methotqic7_2007}. Consequently, it is still unclear what is a good quantifier of nonlocality.

Yet another possibility to quantify the nonlocal correlations of complex states is based on the probability that random measurements generate nonlocal statistics. The {\em probability of violation} of local realism under random measurements, proposed in \cite{Liangprl104_2010,Wallmanpra83_2011}, has gained considerable attention as an operational measure of nonclassicality of quantum states \cite{Fonsecapra92_2015}. It has been demonstrated both numerically \cite{Rosierpra96_2017,Barasinskipra98_2018,Fonsecapra98_2018,Barasinskipra101_2020} and analytically \cite{Fonsecapra92_2015,Lipinskanpj_2018} that this quantity is a good candidate for a nonlocality measure. Furthermore, in \cite{Lipinskanpj_2018} it was proved that this quantifier satisfies some natural properties and expectations for an operational measure of nonclassicality, e.g., invariance under local unitaries. The probability of violation is also often called {\em volume of violation} \cite{Fonsecapra92_2015} or  {\em nonlocal fraction} \cite{Lipinskanpj_2018,Barasinskipra101_2020}. In this paper we will use the latter name. 

The nonlocal fraction of the state $\rho$ is defined as \cite{Liangprl104_2010,Wallmanpra83_2011}
\begin{equation}
\mathcal{P}_V(\rho) = \int f(\rho, \Omega) d \Omega,
\label{eq:pV}
\end{equation}
where we integrate over a space of measurement parameters $\Omega$ that can be varied within a Bell scenario according to the Haar measure and the function $f(\rho, \Omega)$ takes one out of two possible values
\begin{displaymath}
f(\rho, \Omega) = \left\{ \begin{array}{ll}
1 & \textrm{if settings lead to violations}\\
 & \textrm{of local realism,}\\
0, & \textrm{otherwise.}
\end{array} \right.
\end{displaymath}
What is important, in this approach the nonlocal correlations are quantified without any  prior assumptions about specific Bell inequalities. 

Although definition in Eq. \eqref{eq:pV} fairly captures the nonlocality extent of a state, the nonlocal fraction does not provide much information about the strength of nonlocality. Therefore, it seems useful to put it together with another quantitative description, called nonlocality strength, which addresses the “fragility” of this nonlocality against noise \cite{Rosierpra101_2020}.

While the study of the nonlocal fraction and the nonlocality strength offers a promising insight into the geometry of the set of quantum correlations, several crucial aspects towards experimental investigations were not addressed so far in the literature. The main limitation of these quantifiers is that the analysis of them requires a complete knowledge about the Bell local polytope (e.g. a complete set of tight Bell inequalities) to detect violation of local realism. Apart from some of the simplest cases, such complete sets  remain unknown (see \cite{Copepra100_2019} for review). Algorithmically it is possible to resolve the Bell polytope (or alternatively called Bell-Pitowsky polytope), however the problem is NP-hard \cite{Brunnerrmp86_2014}. In general, the number of tight Bell inequalities is expected to grow exponentially with the number of parties, the number of outcomes and the number of measurement settings. 

From a theoretical point of view, the problem can be lifted by applying a linear programming approach and directly considering the space of behaviors (space of joint probabilities), which local polytopes inhabit. In this case, the only context information required being the number of parties, the number of measurements per party and the number of outcomes per measurement \cite{Rosierpra96_2017}. 
Note a recent theoretical work in the multipartite scenario which investigates Bell-nonlocal correlations using linear programming and specific Bell inequalities as well by having the parties performing their measurements in a randomly chosen triad instead of randomly chosen bases~\cite{Yangpra102_2020}.

However, this geometric approach has no direct experimental implementation what causes a lack of experimental studies of the subject. It is because frequencies obtained in measurements are subject to Poissonian statistics, which can lead to the violation of local realism with the state visibility equal to $0$, due to the inability to construct a correct joint probability distribution (see \cite{Grucapra82_2010} for a more detailed discussion). These problems can be avoided by processing the collected experimental data using maximum-likelihood \cite{Hradilpra55_1997} or device-independent point estimation \cite{Linpra97_2018} methods. However, it should be noted that this additionally increases the complexity of the problem, as it requires the analysis of a large amount of experimental data. 
Therefore, the only known experimental results of the nonlocal fraction $\mathcal{P}_V$ \cite{ShadboltSciRep2_2012,Slaternjp16_2014,Barasinskipra101_2020} are related to the Clauser-Horne-Shimony-Holt (CHSH) scenario \cite{CHSHprl23_1969} and Pitowsky-Svozil (PS) scenario \cite{Pitowskypra64_2001,Sliwapla317_2003} as the complete sets of tight Bell inequalities are known in these cases.

A question which naturally arises is whether the nonlocal fraction and the nonlocality strength can be measured experimentally with a partial knowledge about a full set of Bell inequalities. 
So far only a few attempts have been made to solve the problem, showing usually a great underestimation of the exact results. For instance, it was shown in Ref.~\cite{Wallmanpra83_2011} that the nonlocal fraction estimated by means of the Mermin-Ardehali-Belinskii-Klyshko (MABK) and Weinfurter-Werner-Wolf-\.Zukowski-Brukner (WWW\.ZB) inequalities is few times smaller than that of the full set of Bell inequalities. However, it has turned out recently that in some cases, it is possible to indicate an incomplete set of inequalities, which significantly improves the estimation \cite{Barasinskipra101_2020}.
In other words, determining the nonlocality within incomplete set looks promising if choosing Bell's inequality class appropriately. Moreover, it seems that, from an experimental point of view, just such an approach with an incomplete set of Bell inequalities is desirable. By the very definition, $\mathcal{P}_V$ is solely related to the fact of violation of local realism, not to the strength of such violation. Therefore, any Bell inequality which is violated with the strength of being close to the upper threshold, is unsuitable for experimental verification of e.g. the nonlocal fraction, since a violation might be simply accidental due to shot noise. 

In this work, we tackle these problems and analyze the statistical relevance of various classes of Bell inequalities for several Bell scenarios. We show that even in the very general scenario, one such class provides results that are close to those of the full set of Bell inequalities. In other words, one can considerably simplify the procedure towards determining the nonlocal fraction by using only one suitable inequality instead of the complicated linear programming method. Furthermore, the average time complexity in our approach is much smaller that this of linear programming algorithm for the same scenario. Surprisingly, the imprecision arising from this approach is, in general,  of the same magnitude as experimental errors. 
Therefore, these results open a door for experimental verification of many known theoretical predictions for multipartite qubit states. Our predictions were also investigated experimentally for the three-qubit case, showing a good agreement with theoretical results.

\section{Theoretical framework}

In this paper, we consider the multipartite Bell experiment involving $N-$qubit states where $N$ spatially separated observers performing two-outcome measurements on their $N$ local subsystems. Each observer can choose among $m_i$ arbitrary observables, $\{\hat{O}^i_{1},\hat{O}^i_{2},\dots,\hat{O}^i_{m_i}\}$ where ($i = 1,2, . . . ,N$). The observables are defined by orthogonal projections $\hat{O}^i_j=~U^i_j|0\rangle_i\langle 0|-U^i_j|1\rangle_i\langle 1|$, where $U^i_j$ denotes a general unitary transformation belonging to the $U(2)$ group and $|r\rangle_i$ stands for the computational basis state of the $i$th observer. The measurement in each basis provides the observer one out of two possible outcomes, denoted $r^i_j = \{0,1\}$. For simplicity, we will refer to this scenario as $m_1 \times \cdots \times m_N$.

With the above assumption, a local realistic description of a Bell experiment is equivalent to the existence of a joint probability distribution $p_{lr}(r^1_1, \dots,r^1_{m_1}; \dots ;r^N_1, \dots,r^N_{m_N})$, where $r^i_{j_i}$ denotes the result of the measurement performed by the $i$th observer when the observer chooses the $j_i$th measurement setting. If a local model exists, quantum predictions for the probabilities are given by the marginal sums:
\begin{align}
P&(r^1, \cdots, r^N| \hat{O}_{k_1}^1,\cdots, \hat{O}_{k_N}^N)=\textrm{Tr}(\rho \cdot \hat{O}_{k_1}^1 \otimes \cdots \otimes\hat{O}_{k_N}^N)\nonumber\\
&=\sum^{1}_{r^1_{j_1}, \dots, r^1_{j_N}=0} p_{lr}(r^1_1, \dots,r^1_{m_1}, \dots ,r^N_1, \dots,r^N_{m_N}),
\label{eq:LHV}
\end{align}
where $P(r^1, \cdots, r^N|\hat{O}_{k_1}^1,\cdots, \hat{O}_{k_N}^N)$ denotes the probability that all observers simultaneously obtain the respective result $r^i$ while measuring observables $O_{k_i}^i$.
It can be shown that for some quantum entangled states the marginal sums cannot be satisfied, which is an expression of Bell’s theorem. Determining the existence of the local realistic description for a given state and set of observables is a typical linear programming problem~\cite{boyd_vandenberghe_2004}. However, in the case of experimental studies, one should follow a different approach.

In the space of probabilities, the set of local correlations $P(r^1, \cdots, r^N| \hat{O}_{k_1}^1,\cdots, \hat{O}_{k_N}^N)$ which satisfies (\ref{eq:LHV}) (hereafter denoted as $L_N$) is convex with finitely many vertices and called the local polytope \cite{PitowskyLNP_1989}. The $L_N$ polytope is bounded by facets (hyperplanes), which can be described by a linear function of the probabilities 
\begin{align*}
I^{(N)}(\textbf{P})\equiv \sum_{\textbf{r},\hat{\textbf{O}}} w^{\hat{\textbf{O}}}_{\textbf{r}} P(r^1, \cdots, r^N| \hat{O}_{k_1}^1,\cdots, \hat{O}_{k_N}^N) = C_{LHV},
\end{align*} 
where $w^{\hat{\textbf{O}}}_{\textbf{r}}$ and $C_{LHV}$ are real coefficients, and we have simplified the notation by introducing 
\begin{align*}
\textbf{P} =& \{P(r^1, \cdots, r^N| \hat{O}_{k_1}^1,\cdots, \hat{O}_{k_N}^N)\},\\
\hat{\textbf{O}} =& \{\hat{O}_{k_1}^1,\cdots, \hat{O}_{k_N}^N\}, 
\end{align*}
and $\textbf{r} = \{r^1, \cdots, r^N\}$.
Correlations which do not admit the decomposition in Eq.~\eqref{eq:LHV} are referred to as nonlocal and lie outside the local polytope $L_N$. In other words, they must violate at least one inequality $I^{(N)}(\textbf{P}) \leq C_{LHV}$. Such inequalities are called tight Bell inequalities and $C_{LHV}$ depicts the upper threshold of inequality $I^{(N)}$ for local realism. In order to determine the nonlocal fraction $\mathcal{P}_V$ for a given state, we calculate how many sets of settings (in percents) lead to violation of local realism, i.e., whether the decomposition in Eq. \eqref{eq:LHV} exists or alternatively whether all Bell inequalities $I^{(N)}$ for a given Bell scenario are satisfied. As in general, the full set of tight Bell inequalities is unknown for a given Bell scenario, in the rest of the text we emphasize by $\mathcal{P}_V^{L_N}$ the fact (if necessary) that results were obtained with the linear programming method and refer to the whole polytope $L_N$, while $\mathcal{P}_V^{I}$ corresponds to a subset of Bell inequalities.

Usually for experimental purposes, an alternative parametrization of $I^{(N)}(\textbf{P})$ is used. It is based on correlation coefficients, e.g.,  $\langle E^{i}_{k_i} \rangle$, $\langle E^{i}_{k_i} E^{j}_{k_{j}}\rangle$ etc., which satisfy the relation 
\begin{align*}
P(r^1, \cdots, r^N| \hat{O}_{k_1}^1,\cdots, \hat{O}_{k_N}^N) &= \\
\frac{1}{2^N} [1 + \sum_{i=1}^N (-1)^{r^i} \langle E^{i}_{k_i} \rangle
+ \sum_{i=1}^{N-1} &\sum_{j=i+1}^{N}(-1)^{r^i r^j} \langle E^{i}_{k_i} E^{j}_{k_j} \rangle \\
+ \dots +(-1)^{\prod^{N}_{i=1} r^i} &\langle E^{1}_{k_1} \cdots E^{N}_{k_N}\rangle]
\end{align*}
and have a clear experimental interpretation. 
For instance, in an experimental setup based on correlated photons each correlation coefficient can be expressed as a function of coincidence counts measured on the detectors \cite{Barasinskipra98_2018}. 

The degree of violation of the Bell inequality $I^{(N)}(\textbf{P})$ is also directly related with the so-called resistance to noise i.e. the amount of white noise admixture required to completely suppress the nonlocal character of the original correlations of a given state $\rho$. Specifically, if for the state $\rho$ and particular choice of measurement settings $\langle I^{(N)}(\textbf{P}) \rangle > C_{LHV}$ then a new state $\sigma(v) = v \rho + (1-v) \openone/2^N$ also reveals nonlocality for $v \geq v_{crit}$, where $v_{crit}$ is a critical value of $v$, for which $C_{LHV} = \langle I^{(N)}(\textbf{P})\rangle $.

Following this observation, a new quantity called nonlocality strength, $\mathcal{S}$, can be defined \cite{Rosierpra101_2020}. It is given by $\mathcal{S}=1-v_{crit}$. Furthermore, it is convenient to use the average value of nonlocality strength:
\begin{equation}
\bar{\mathcal{S}}(\rho) = \int_0^{\mathcal{S}^{\max}} \mathcal{S} g(\rho,\mathcal{S}) d\mathcal{S},
\end{equation}
where $g(\rho,\mathcal{S})$ is a nonlocality strength distribution and $\mathcal{S}^{\max}$ depict a highest attainable nonlocal strength with respect to the full set of tight Bell inequalities and measurement settings. The nonlocality strength distribution is given by
\begin{equation}
g(\rho,\mathcal{S}) = \int \omega(\rho, \mathcal{S}, \Omega) d \Omega,
\label{eq:gDis}
\end{equation}
where we integrate over a space of measurement parameters $\Omega$ that can be varied within a Bell scenario according to the Haar measure and the function $\omega(\rho, \mathcal{S}, \Omega)$ takes one out of two possible values
\begin{displaymath}
\omega(\rho, \mathcal{S}, \Omega) = \left\{ \begin{array}{ll}
1 & \textrm{if settings lead to nonlocal}\\
 & \textrm{strength in range $(\mathcal{S},\mathcal{S}+d\mathcal{S})$,}\\
0, & \textrm{otherwise.}
\end{array} \right.
\end{displaymath}
Our results are normalized such that the areas of the regions bounded by the plots directly provide the nonlocal fraction, $\int_0^{\mathcal{S}^{\max}} g(\mathcal{S}) d\mathcal{S} = \mathcal{P}_V$ \cite{Rosierpra101_2020}.

\begin{figure}
\centering
\includegraphics[width=\columnwidth]{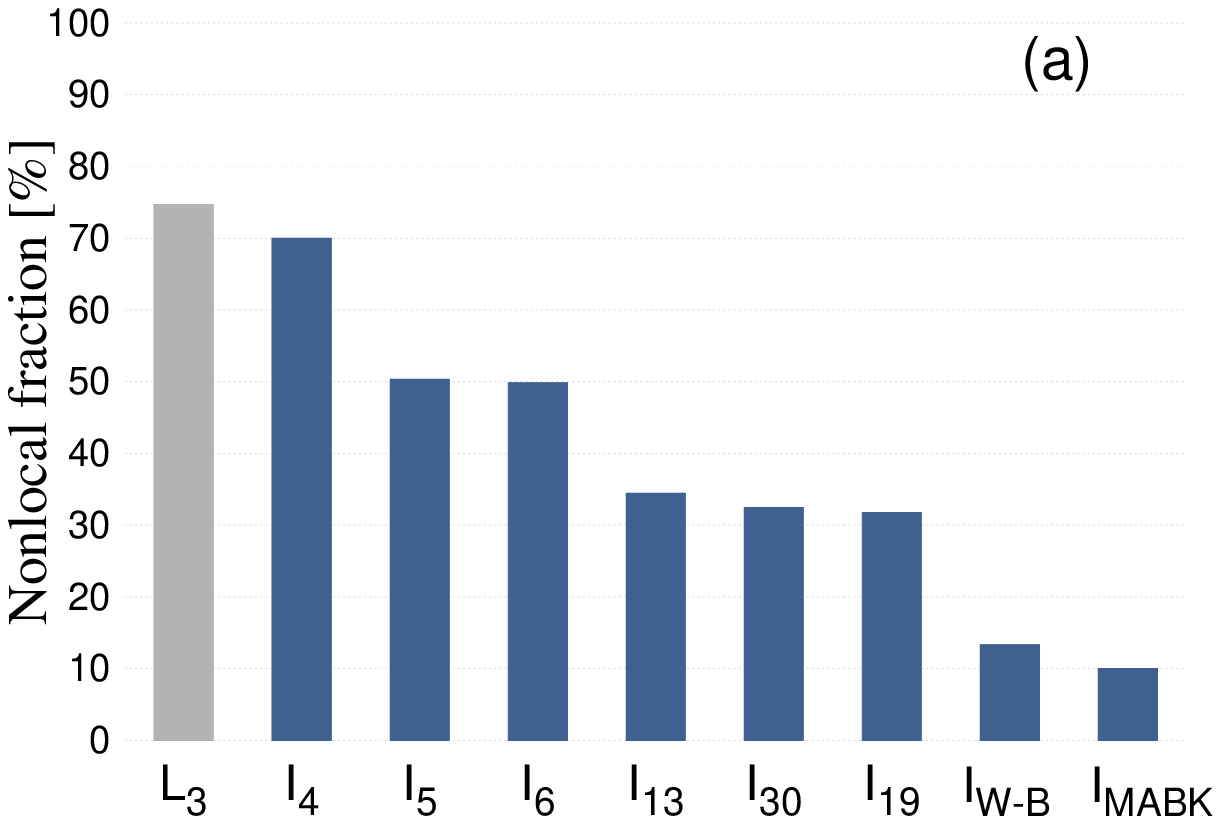}
\includegraphics[width=\columnwidth]{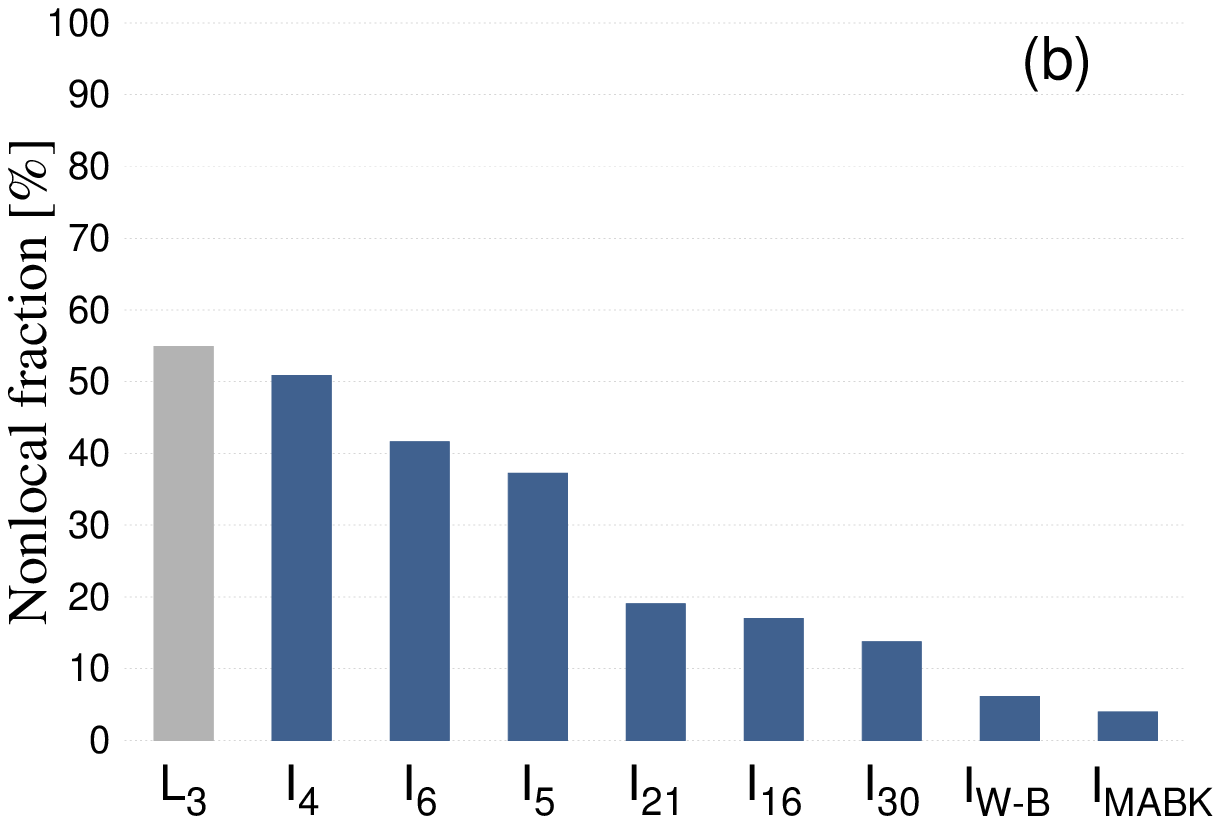}
\caption{Nonlocal fraction calculated for (a) the three-qubit GHZ state and (b) the three-qubit W state. Gray bar represents results for the polytope $L_3$ while blue bars correspond to outcomes for various Bell inequalities presented in Appendix \ref{appendixA}. Note that $I_{\text{W-B}}$ stands for $I_{\text{WWW\.ZB}}$.
}
\label{fig:L3a}
\end{figure}

\section{Numerical Results}
\subsection{Statistical relevance of Bell inequalities}

In order to present the main result, let us consider a family of Bell inequalities (equivalent under permutation of parties, inputs, and outputs) which can be obtained in a lifting procedure applied to the CHSH inequalities \cite{Pironiojmp46_2005}. The family is given by
\begin{eqnarray}
I^{(N)}_{opt} = \langle (I^{(2)}_{opt}-2) \prod^{N}_{i=3}(1-E^{(i)}_0)\rangle \leq 0,
\label{eq:I_opt}
\end{eqnarray}
where $I^{(2)}_{opt} \equiv E^{(1)}_{0}E^{(2)}_{0}+E^{(1)}_{1}E^{(2)}_{0}+E^{(1)}_{0}E^{(2)}_{1}-E^{(1)}_{1}E^{(2)}_{1}$ stands for the CHSH expression \cite{CHSHprl23_1969} and $E^{(i)}_j$ denotes an observable measured by the $i$th observer when the $j$th measurement setting is chosen. In other words, the inequality $I^{(N)}_{opt}$ is composed of $2$ observers operating the CHSH experiment and $(N-2)$ observers performing only a single measurement $E^{(i)}_j$. As the $(N-2)$ single measurements $E^{(i)}_j$ cannot cause any violation of local realism, it implies that the nonlocal correlations witnessed by $I^{(N)}_{opt}$ have a two-qubit CHSH origin. It was also shown that $I^{(N)}_{opt}$ is a tight Bell inequality, violated by all pure entangled states of a given number of parties $N$ \cite{Laskowskijpa48_2015}. The inequality also allows us to observe non-classical correlations of highly noisy states \cite{Laskowskijpa48_2015} and with ineffective detectors \cite{Kostrzewapra98_2018}. Moreover, it finds useful applications in witnessing entanglement depth~\cite{Curchodpra91_2015} and selft-testing of multipartite quantum states~\cite{Jebarathinamprr1_2019}.

We argue that these family of Bell inequalities is sufficient to estimate the nonlocal fraction and the nonlocality strength of $N$-qubit states. As there is no proof of the optimality of $I^{(N)}_{opt}$, our working conjecture is supported by the numerical results presented below.

It is also important to stress that for the $m-$setting scenario (i.e. $m_i =m$ for $i=1,2, . . . ,N$) the number of equivalent inequalities $I^{(N)}_{opt}$ is given by $\mathcal{R} =m^2(m-1)^2 N(N-1) 2^{m(N-2)}$. Therefore, the overall time complexity (where we count the number of arithmetic operations) is given by $T_1 = O(\mathcal{\mathcal{R}})=O(m^4 N^2 2^{m(N-2)})$.
On the other hand, using linear programming method in the average time complexity case, we have at least $T_2=O(m^{2N} 2^{m N})$ \cite{Matousek16_1996}. Comparing these two quantities, one can easily notice that $T_1$ is definitely smaller than $T_2$ for $N\geq2$. In other words, it is less costly to generate all the different liftings of the CHSH inequality than running the linear programming algorithm for the same scenario.

\begin{table}[h!]
\caption{Nonlocal fraction $\mathcal{P}_V$ and average nonlocality strength $\bar{\mathcal{S}}$ for the $|{\rm GHZ}_3\rangle$ and $|{\rm W}_3\rangle$ states. The symbol $L_3$ denotes results obtained with linear programming method for $2 \times 2 \times 2$ measurement settings \cite{Rosierpra96_2017}. The abbreviation $Sym.$ stands for the number of equivalent Bell inequalities of a given family.Note that $I_{\text{W-B}}$ stands for $I_{\text{WWW\.ZB}}$.
}
\label{tab:tab_222_results}
%\begin{ruledtabular}
%\resizebox{\colomnwidth}{!} {
\renewcommand*{\arraystretch}{1.13}
\begin{tabular}{@{}c@{\hspace{3pt}}c@{\hspace{6pt}}c@{\hspace{6pt}}c@{\hspace{7pt}}c@{\hspace{6pt}}c@{\hspace{6pt}}c@{}}\toprule
 & & & \multicolumn{2}{c}{$GHZ_3$} & \multicolumn{2}{c}{$W_3$} \\
\cmidrule(r){4-5}
\cmidrule(r){6-7}
Ineq. & Settings & Sym. & $\mathcal{P}_V[\%]$ & $\bar{\mathcal{S}}$ & $\mathcal{P}_V[\%]$ & $\bar{\mathcal{S}}$ \\
\hline\hline
$L_3$ & $2 \times 2\times 2$ &  & $74.688$ & $0.0881$ & $54.893$ & $0.0610$ \\ \hline
$I^{(3)}_{opt}$	& $2 \times 2\times 2$ & $96$ & $69.997$ & $0.0782$ & $50.858$ & $0.0574$ \\
$I_{5}$ & $2 \times 2\times 2$ & $512$ & $50.310$ & $0.0403$ & $37.221$ & $0.0298$ \\
$I_{6}$ & $2 \times 2\times 2$ & $1536$ & $49.858$ & $0.0383$ & $41.617$ & $0.0332$ \\
$I_{13}$ & $2 \times 2\times 2$ & $384$ & $34.426$ & $0.0253$ & $5.252$ & $0.0026$ \\
$I_{19}$ & $2 \times 2\times 2$ & $1536$ & $31.754$ & $0.0220$ & $9.448$ & $0.0045$ \\
$I_{21}$ & $2 \times 2\times 2$ & $1536$ & $20.534$ & $0.0108$ & $19.014$ & $0.0113$ \\
$I_{16}$ & $2 \times 2\times 2$ & $1536$ & $25.878$ & $0.0179$ & $16.988$ & $0.0106$ \\
$I_{30}$ & $2 \times 2\times 2$ & $3072$ & $32.445$ & $0.0220$ & $13.714$ & $0.0082$ \\
$I_{\text{W-B}}$ & $2 \times 2\times 2$ & $1$ & $13.313$ & $0.0149$ & $6.105$ & $0.0043$ \\
$I_{\text{MABK}}$ & $2 \times 2\times 2$ & $16$ & $10.002$ & $0.0123$ & $4.835$ & $0.0038$ \\
\bottomrule
\end{tabular}
%}
%\end{ruledtabular}
\end{table}

\subsubsection{$2 \times 2 \times 2$ Bell scenario}

Let us start with the three spatially separated observers performing one out of two dichotomic measurements, i.e. the $2 \times 2 \times 2$ case. This scenario is completely characterized by $46$ classes (families) of Bell inequalities derived by Pitowsky and Svozil \cite{Pitowskypra64_2001}. One of such classes (namely, the $4th$ facet inequality) is given by Eq. \eqref{eq:I_opt}. First, we determine the statistical relevance of these $46$ classes concerning the nonlocal fraction. We calculate $\mathcal{P}_V$ of each class independently for two inequivalent types of tripartite entangled states, namely the ${\rm GHZ}_3$ and ${\rm W}_3$ state \cite{Durpra62_2000} defined in the~Appendix~\ref{appendixB}. 
Our results for the most relevant cases are presented in Fig. \ref{fig:L3a}, where the families of Bell inequalities are numbered in the same manner as in Ref. \cite{Sliwapla317_2003} and listed in Appendix \ref{appendixA}: Table \ref{tab:tab_222}. 

As we see, depending on the chosen entangled state the statistical relevance of individual families may vary but the best three items remain unchanged. They are given by the class of $4th$, $5th$, and $6th$ facet inequality \cite{Sliwapla317_2003} (hereafter $I^{(3)}_{opt}, I_5$, and $I_6$, respectively). Moreover, apart from these three classes, the nonlocal fraction for other families of Bell inequalities do not exceed $\frac{1}{2} \mathcal{P}_V^{L_3}$ neither for the ${\rm GHZ}_3$ nor for the ${\rm W}_3$ state (see Table \ref{tab:tab_222_results}). In particular, the MABK and WWW\.ZB inequalities, previously discussed in Refs. \cite{Liangprl104_2010,Wallmanpra83_2011}, provide results much smaller than these of the best three items. Interestingly, even if we consider such $43$ classes simultaneously (the complete set of Bell inequalities, excluding $I^{(3)}_{opt}$, $I_5$, and $I_6$), the resulting nonlocal fraction is not greater than $57\%$ and $30\%$ for the ${\rm GHZ}_3$ and ${\rm W}_3$ state, respectively. To highlight this phenomenon, we note that these $43$ classes contain $51712$ tight Bell inequalities while $I^{(3)}_{opt}$, $I_5$, and $I_6$ corresponds only to $96$, $512$, and $1536$ inequalities, respectively (see Table \ref{tab:tab_222_results}). Therefore, all the upper-mentioned $43$ classes are rather unsuitable for a potential experimental measure of the nonlocal fraction.

\begin{figure}
\centering
\includegraphics[width=\columnwidth]{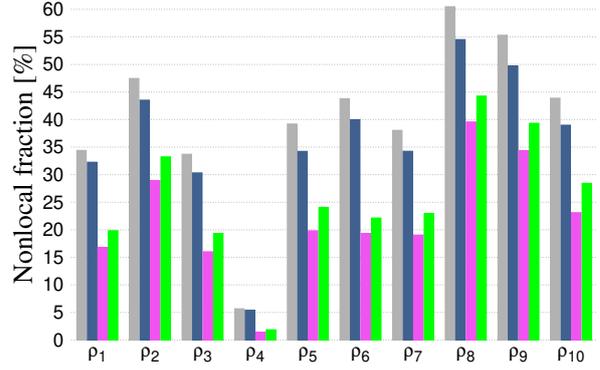}
\caption{Nonlocal fraction calculated for randomly generated three-qubit states $\rho_i$. For each state the gray bar represents results for the polytope $L_3$ while the other three bars correspond to outcomes for inequalities $I^{(3)}_{opt}, I_5$, and $I_6$, respectively. Note that these three Bell inequalities also provide the highest three values of $\mathcal{P}_V^{I_{j}}(\rho_i)$.}
\label{fig:S30}
\end{figure}

Among the $I^{(3)}_{opt}$, $I_5$, and $I_6$ families, a clear dominant position is reserved for the $4th$ facet inequality. As shown in Table \ref{tab:tab_222_results}, this family provides a very good approximation of $\mathcal{P}_V^{L_3}$ with a gap smaller than $5.3~p.p.$, i.e.  $\mathcal{P}_V^{L_3}({\rm GHZ}_3) = 1.031 \mathcal{P}_V^{I_{opt}}({\rm GHZ}_3)$ and $\mathcal{P}_V^{L_3}({\rm W}_3) = 1.053 \mathcal{P}_V^{I_{opt}}({\rm W}_3)$. Furthermore, the $I^{(3)}_{opt}$ family can also be used to estimate the nonlocal fraction for other kinds of states, like the generalized nonmaximally entangled ${\rm GHZ}_3$ states (see Ref. \cite{Barasinskipra101_2020}).  In order to highlight this observation, we have examined $100$ randomly generated pure states. As a result, we have found that the $I^{(3)}_{opt}$, $I_5$, and $I_6$ families are also the dominant ones and the best approximation of $\mathcal{P}_V^{L_3}$ is provided by the $I^{(3)}_{opt}$ family (see Fig. \ref{fig:S30}).

\begin{figure}
\centering
\includegraphics[width=\columnwidth]{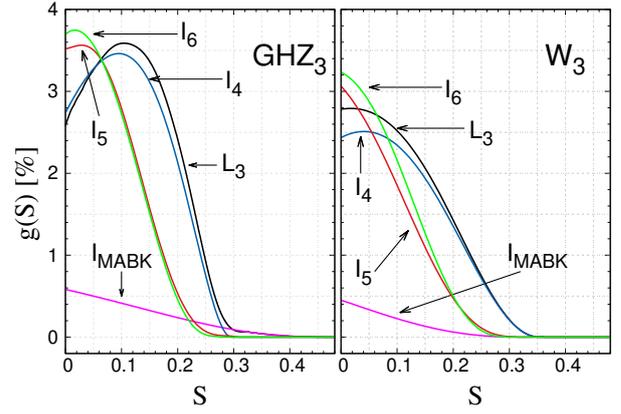}
\caption{Nonlocality strength distributions for three-qubit states with $2 \times 2 \times 2$ measurements settings. The symbol $L_3$ denotes calculation made for the whole polytope while $I_n$ depicts predictions for respective Bell inequality}
\label{fig:S3a}
\end{figure}

Analogous conclusion can be drawn when the nonlocality strength, $\mathcal{S}$, is  taken into consideration. Once again we see (Table~\ref{tab:tab_222_results}) that the best  approximation of the average strength, $\bar{\mathcal{S}}$, is provided by the $I^{(3)}_{opt}$ family,  what suggests a similar shape of the nonlocality strength distribution for both cases, $L_3$ and $I^{(3)}_{opt}$. Intuitively, one may expect such outcome, assuming that the better estimation of nonlocal fraction, the closer approximation of $g(\mathcal{S})$. Although that assumption is correct for the cases in question (see Fig \ref{fig:S3a}), in general, it is not true as will be discussed later. Despite of that another very interesting remarks towards an experimental implementation can be made on the distribution $g(\mathcal{S})$. As we see in Fig.~ \ref{fig:S3a}, the nonlocality strength distribution for $I^{(3)}_{opt}$ has a similar shape as $g(\mathcal{S})$ for the politope $L_3$. Any differences are either small in magnitude or appear for small $\mathcal{S}$. For instance, the function $g^{I_{opt}}(\mathcal{S})$ for ${\rm GHZ}_3$ state vanishes above $\mathcal{S}>0.29$ what stands in contrast to $g^{L_3}(\mathcal{S})$, due to the strong violation of MABK inequality. On the other hand, the greatest difference between $g^{I_{opt}}(\mathcal{S})$ and $g^{L_3}(\mathcal{S})$ for W state is observed for $\mathcal{S}<0.1$. Bearing in mind that an experimental detection of the nonlocality (violation of Bell inequality) is ambiguous when the nonlocal strength is close to the measurement accuracy (say $\mathcal{S}=\pm 0.015$ \cite{Barasinskipra99_2019}), this tendency should imply a positive impact on the potential experimental results by decreasing experimental errors. The classes of the $5th$ and $6th$ facet inequality, on the other hand, indicate an opposite trend and they overestimate $g^{L_3}(\mathcal{S})$ in the regime of small $\mathcal{S}$ while strongly underestimate $g^{L_3}(\mathcal{S})$ for $\mathcal{S}>0.06$ (Fig~\ref{fig:S3a}).

\begin{table}
\caption{Nonlocal fraction $\mathcal{P}_V$ and average nonlocality strength $\bar{\mathcal{S}}$ for the $|{\rm GHZ}_3\rangle$ and $|{\rm W}_3\rangle$ states. The symbol $L_3^{ijk}$ corresponds to results in Ref. \cite{Rosierpra96_2017}.
}
\label{tab:tab_322_results}
%\begin{ruledtabular}
\renewcommand*{\arraystretch}{1.13}
\begin{tabular}{@{}c@{\hspace{3pt}}c@{\hspace{6pt}}c@{\hspace{6pt}}c@{\hspace{7pt}}c@{\hspace{6pt}}c@{\hspace{6pt}}c@{}}\toprule
 & & & \multicolumn{2}{c}{${\rm GHZ}_3$} & \multicolumn{2}{c}{${\rm W}_3$} \\
\cmidrule(r){4-5}
\cmidrule(r){6-7}
Ineq. & Settings & Sym. & $\mathcal{P}_V[\%]$ & $\bar{\mathcal{S}}$ & $\mathcal{P}_V[\%]$ & $\bar{\mathcal{S}}$ \\
\hline\hline
$L^{322}_3$ & $3 \times 2\times 2$ & & $90.132$ & $0.1326$ & $76.788$ & $0.1011$ \\ \hline
$I^{(3)}_{opt}$ & $3 \times 2\times 2$ & $240$ & $87.415$ & $0.1176$ & $72.931$ & $0.0962$\\
$I_5$ & $3 \times 2\times 2$ & $1536$ & $73.279$ & $0.0698$ & $60.291$ & $0.0555$\\
$I_6$ & $3 \times 2\times 2$ & $4608$ & $74.337$ & $0.0685$ & $64.757$ & $0.0605$\\
$I^{322}_1$ & $3 \times 2\times 2$ & $192$ & $18.242$ & $0.0216$ & $11.538$ & $0.0085$\\
$I^{322}_2$ & $3 \times 2\times 2$ & $768$ & $14.343$ & $0.0083$ & $2.394$ & $0.0010$\\
$I^{322}_3$ & $3 \times 2\times 2$ & $1536$ & $36.920$ & $0.0260$ & $12.907$ & $0.0063$\\
$I^{322}_4$ & $3 \times 2\times 2$ & $3072$ & $58.577$ & $0.0645$ & $24.445$ & $0.0172$\\
$I^{322}_5$ & $3 \times 2\times 2$ & $3072$ & $40.624$ & $0.0266$ & $30.394$ & $0.0169$\\
$I^{322}_6$ & $3 \times 2\times 2$ & $384$ & $33.749$ & $0.0247$ & $6.575$ & $0.0030$\\
$I^{322}_7$ & $3 \times 2\times 2$ & $192$ & $36.127$ & $0.0346$ & $6.467$ & $0.0034$\\
$I^{322}_8$ & $3 \times 2\times 2$ & $3072$ & $35.061$ & $0.0232$ & $33.108$ & $0.0234$\\
$I^{322}_9$ & $3 \times 2\times 2$ & $768$ & $33.303$ & $0.0244$ & $11.182$ & $0.0057$\\
\hline\hline
$L^{332}_3$ & $3 \times 3\times 2$ & & $97.245$ & $0.1781$ & $91.366$ & $0.1464$ \\ \hline
$I^{(3)}_{opt}$ & $3 \times 3\times 2$ & $576$ & $96.109$ & $0.1564$ & $88.893$ & $0.1413$\\
$I_5$ & $3 \times 3\times 2$ & $4608$ & $89.907$ & $0.1060$ & $81.867$ & $0.0921$\\
$I_6$ & $3 \times 3\times 2$ & $13824$ & $90.591$ & $0.1048$ & $84.461$ & $0.0975$\\
%$I^{322}_1$ & $3 \times 3\times 2$ & $33.604$ & $0.0456$ & $26.737$ & $0.0226$\\
%$I^{322}_2$ & $3 \times 3\times 2$ & $40.259$ & $0.0267$ & $9.061$ & $0.0042$\\
$I^{322}_3$ & $3 \times 3\times 2$ & $9216$ & $75.091$ & $0.0702$ & $28.067$ & $0.0210$\\
$I^{322}_4$ & $3 \times 3\times 2$ & $18432$ & $77.031$ & $0.0759$ & $46.503$ & $0.0368$\\
$I^{322}_5$ & $3 \times 3\times 2$ & $18432$ & $76.427$ & $0.0672$ & $62.228$ & $0.0439$\\
$I^{322}_6$ & $3 \times 3\times 2$ & $2304$ & $71.224$ & $0.0681$ & $20.233$ & $0.0103$\\
$I^{322}_7$ & $3 \times 3\times 2$ & $1152$ & $74.913$ & $0.0940$ & $21.218$ & $0.0125$\\
$I^{322}_8$ & $3 \times 3\times 2$ & $18432$ & $75.338$ & $0.0874$ & $31.291$ & $0.0202$\\
$I^{322}_9$ & $3 \times 3\times 2$ & $4608$ & $72.019$ & $0.0696$ & $31.850$ & $0.0187$\\
$I^{332}_1$ & $3 \times 3\times 2$ & $1152$ & $26.016$ & $0.0299$ & $9.865$ & $0.0067$\\
$I^{332}_2$ & $3 \times 3\times 2$ & $2304$ & $30.800$ & $0.0376$ & $36.399$ & $0.0223$\\
$I^{332}_3$ & $3 \times 3\times 2$ & $4608$ & $60.269$ & $0.0693$ & $38.840$ & $0.0276$\\
$I^{332}_4$ & $3 \times 3\times 2$ & $2304$ & $60.383$ & $0.0717$ & $33.067$ & $0.0236$\\
$I^{332}_5$ & $3 \times 3\times 2$ & $9216$ & $39.399$ & $0.0377$ & $6.933$ & $0.0031$\\
$I^{332}_6$ & $3 \times 3\times 2$ & $1152$ & $40.161$ & $0.0384$ & $28.342$ & $0.0194$\\
\hline\hline
$L^{333}_3$ & $3 \times 3\times 3$ & & $99.542$ & $0.2228$ & $97.797$ & $0.1901$ \\ \hline
$I^{(3)}_{opt}$ & $3 \times 3\times 3$ & $1296$ & $99.217$ & $0.1904$ & $96.698$ & $0.1846$\\
$I_5$ & $3 \times 3\times 3$ & $13824$ & $97.635$ & $0.1436$ & $94.601$ & $0.1345$\\
$I_6$ & $3 \times 3\times 3$ & $41472$ & $97.807$ & $0.1420$ & $95.447$ & $0.1386$\\
$I^{322}_3$ & $3 \times 3\times 3$ & $41472$ & $94.394$ & $0.1206$ & $64.761$ & $0.0468$\\
$I^{322}_4$ & $3 \times 3\times 3$ & $82944$ & $99.063$ & $0.1985$ & $86.470$ & $0.0963$\\
$I^{322}_5$ & $3 \times 3\times 3$ & $82944$ & $94.041$ & $0.1122$ & $85.611$ & $0.0790$\\
$I^{322}_7$ & $3 \times 3\times 3$ & $5184$ & $95.408$ & $0.1614$ & $44.581$ & $0.0307$\\
$I^{322}_8$ & $3 \times 3\times 3$ & $82944$ & $91.729$ & $0.1061$ & $87.194$ & $0.1003$\\
$I^{332}_2$ & $3 \times 3\times 3$ & $20736$ & $57.742$ & $0.0744$ & $33.501$ & $0.0327$\\
$I^{332}_3$ & $3 \times 3\times 3$ & $41472$ & $95.307$ & $0.1607$ & $72.950$ & $0.0696$\\
$I^{332}_4$ & $3 \times 3\times 3$ & $20736$ & $95.562$ & $0.1669$ & $66.971$ & $0.0630$\\
$I^{332}_5$ & $3 \times 3\times 3$ & $82944$ & $85.890$ & $0.1120$ & $29.457$ & $0.0156$\\
$I^{332}_6$ & $3 \times 3\times 3$ & $10368$ & $89.909$ & $0.0760$ & $67.152$ & $0.0603$\\
$I^{333}_1$ & $3 \times 3\times 3$ & $6912$ & $39.570$ & $0.0505$ & $46.906$ & $0.0423$\\ 
$I^{333}_2$ & $3 \times 3\times 3$ & $20736$ & $49.665$ & $0.0630$ & $48.946$ & $0.0486$\\
%\end{eqnarray}
\bottomrule
\end{tabular}
%\end{ruledtabular}
\end{table}

\subsubsection{$m_1 \times m_2 \times m_3$ Bell scenario}

Next, we analyze the scenario when the number of measurement settings increases. In particular, we extend our studies to the $3 \times 2 \times 2$, $3 \times 3 \times 2$, and $3 \times 3 \times 3$ scenario. As a complete set of tight Bell inequalities for any of these cases is unknown, we employed a linear programming method \cite{boyd_vandenberghe_2004} to identify an explicit form of the most relevant Bell inequalities, which were the most frequently violated for given (random) measurement settings. 
Some of them naturally overlap with the inequalities derived by Pitowsky and Svozil \cite{Pitowskypra64_2001} but genuine $m_1 \times m_2 \times m_3$ inequalities also belong to that group. All identified genuine inequalities are listed in Appendix \ref{appendixA}: Table \ref{tab:tab_322}. Every such expression represents a distinct class of Bell inequalities, equivalent under permutation of parties, inputs, and outputs \cite{Wernerpra64_2001,Collins_2004}. 

\begin{figure}
\centering
\includegraphics[width=\columnwidth]{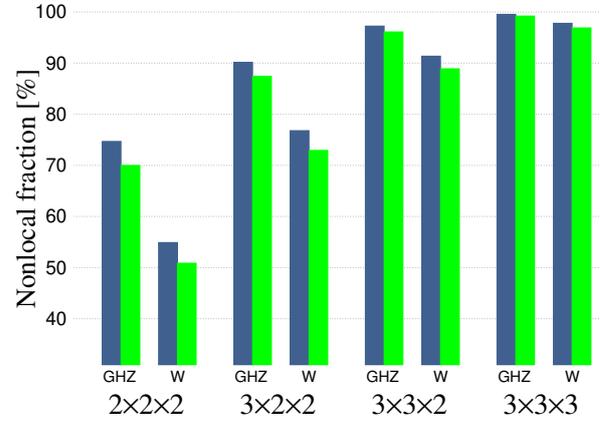}
\caption{Comparison between the nonlocal fraction calculated with linear programming method \cite{Rosierpra96_2017} (blue bars) and the nonlocal fraction estimated by mean $I^{(3)}_{opt}$ (green bars).}
\label{fig:fig4}
\end{figure}

Based on our identification, the nonlocal fraction and the average strength of nonlocality has been calculated for each $m_1 \times m_2 \times m_3$ scenarios mentioned above. The results are collected in Table \ref{tab:tab_322_results} from which the following remarks can be drawn: 

(i) As we see, for each scenario the highest $\mathcal{P}_V$ and $\bar{\mathcal{S}}$ is always achieved for the $I^{(3)}_{opt}$ family. Moreover, the gap between $\mathcal{P}_V^{L_3}$ and $\mathcal{P}_V^{I_{opt}}$  decreases with the number of measurement settings (see Fig. \ref{fig:fig4}). In other words, for three observers the $I^{(3)}_{opt}$ family seems to be sufficient tool for experimental determination of $\mathcal{P}_V$. 

(ii) As shown in Ref. \cite{Rosierpra96_2017,Lipinskanpj_2018} the nonlocal fraction increases rapidly when the number of measurement settings grows. This property might arise either by the increase of the number of equivalent Bell inequalities or by the emergence of new types of Bell inequalities. Our results and our previous remark imply that the increase of nonlocal fraction with $m_i$ has a statistical explanation (at least for three-qubit states), i.e. by increasing the number of settings, we increase the number of equivalent Bell inequalities that belong to the $I^{(3)}_{opt}$ family and hence, the probability that some of them are violated, involving only two settings. Despite such limitations, the estimation of $\mathcal{P}_V$ is completely consistent with the previous calculations \cite{Rosierpra96_2017}. It is worth emphasizing that the genuine $m_1 \times m_2 \times m_3$ inequalities (e.g $I^{322}_4$, $I^{332}_4$ etc.) provide the nonlocal fraction significantly smaller than that of the $I^{(3)}_{opt}$ family. 

(iii) In the case of $\bar{\mathcal{S}}$, an increase of the number of measurement settings implies a growth of the gap between $\bar{\mathcal{S}}^{L_3}$ and $\bar{\mathcal{S}}^{I_{opt}}$, although very slight if the $W_3$ states are taken under consideration. This behavior is caused by a considerable reduction in the number of Bell inequalities describing local polytopes and will be further discussed in the next section.

\begin{table*}
	\caption{Nonlocal fraction $\mathcal{P}_V$ and average nonlocality strength $\bar{\mathcal{S}}$ for various four-qubit states and $I^{(4)}_{opt}$ inequality.
	}
	\label{tab:tab4qubity}
%	\begin{ruledtabular}
\renewcommand*{\arraystretch}{1.13}
		\begin{tabular}{@{}c@{\hspace{12pt}} c@{\hspace{12pt}} c@{\hspace{11pt}} c@{\hspace{12pt}} c@{\hspace{11pt}} c@{\hspace{12pt}} c@{\hspace{11pt}} c@{\hspace{12pt}} c@{\hspace{11pt}} c@{\hspace{12pt}} c@{\hspace{11pt}} c@{}}\toprule
			& & \multicolumn{2}{c}{${\rm GHZ}_2 \otimes |00\rangle$} & \multicolumn{2}{c}{${\rm GHZ}_4$} & \multicolumn{2}{c}{${\rm W}_4$} & \multicolumn{2}{c}{${\rm Cl}_4$} & \multicolumn{2}{c}{${\rm D}^2_4$} \\
			\cmidrule(r){3-4}
			\cmidrule(r){5-6}
			\cmidrule(r){7-8}
			\cmidrule(r){9-10}
			\cmidrule{11-12}
			Settings & Sym. & $\mathcal{P}_V[\%]$ & $\bar{\mathcal{S}}$ & $\mathcal{P}_V[\%]$ & $\bar{\mathcal{S}}$ & $\mathcal{P}_V[\%]$ & $\bar{\mathcal{S}}$ & $\mathcal{P}_V[\%]$ & $\bar{\mathcal{S}}$ & $\mathcal{P}_V[\%]$ & $\bar{\mathcal{S}}$ \\ 
			\hline\hline
			$2 \times 2 \times 1 \times 1$ & $32$ & $28.318$ & $0.0536$ & $18.744$ & $0.0137$ & $13.260$ & $0.0120$ & $43.047$ & $0.0483$ & $9.452$ & $0.0064$ \\ 
			$2 \times 2 \times 2 \times 1$ & $192$ & $28.318$ & $0.0580$ & $58.049$ & $0.0509$ & $47.563$ & $0.0499$ & $73.909$ & $0.0913$ & $37.017$ & $0.0279$ \\
			$2 \times 2 \times 2 \times 2$ & $768$ & $28.318$ & $0.0626$ & $88.562$ & $0.1067$ & $81.522$ & $0.1134$ & $95.982$ & $0.1618$ & $73.420$ & $0.0701$ \\
			$3 \times 2 \times 2 \times 2$ & $1728$ & $52.401$ & $0.1299$ & $96.400$ & $0.1376$ & $92.987$ & $0.1573$ & $98.834$ & $0.1984$ & $88.139$ & $0.1013$ \\
			$3 \times 3 \times 2 \times 2$ & $3744$ & $78.219$ & $0.2315$ & $98.940$ & $0.1650$ & $97.999$ & $0.1993$ & $99.779$ & $0.2369$ & $95.908$ & $0.1332$ \\
			$3 \times 3 \times 3 \times 2$ & $7776$ & $78.219$ & $0.2394$ & $99.723$ & $0.1884$ & $99.567$ & $0.2367$ & $99.948$ & $0.2594$ & $98.940$ & $0.1631$ \\
		\bottomrule
		\end{tabular}
%	\end{ruledtabular}
\end{table*}

\subsubsection{General Bell scenario involving $N$-qubit state}

Let us now investigate the statistical aspects of nonlocal correlations for several four- and five-qubit states, using only the Bell inequality $I^{(N)}_{opt}$. As for $N-$parties systems, numerous inequivalent kinds of entanglement exist, in this subsection we investigate the behavior of some archetypal four- and five-qubit maximally entangled states. They are explicitly defined in the~Appendix~\ref{appendixB}.

We started by discussing the case in which the quantum state under study is a product of two states, $|\psi\rangle = |{\rm GHZ}_2\rangle\otimes|00\rangle$, what clearly illustrates the nature of the $I^{(N)}_{opt}$ inequality. In this case, the nonlocal fraction takes one of several recurring values, depending on the number of measurement settings $m_1$ and/or $m_2$, e.g. $28.318\%$, $52.401\%$, and $78.219\%$ as in Table \ref{tab:tab4qubity}. The number of measurement settings $m_3$ and $m_4$ is irrelevant and has no influence on $\mathcal{P}_V$. It is because the second part of the analyzed state cannot reveal any kind of nonlocality, regardless of the projectors $E^{(i)}_j$. The only valuable projection of the state $|\psi\rangle$ onto two-qubit state is the one which gives maximally entangled state $|{\rm GHZ}_2\rangle$. Therefore, the nonlocal fraction of $|\psi\rangle$ coincides exactly with the corresponding results of $|{\rm GHZ}_2\rangle$ (see Ref. \cite{Rosierpra96_2017}). 
However, the value of $m_3$ and $m_4$ undoubtedly influences the resulting nonlocal strength (see Table \ref{tab:tab4qubity}). Furthermore, the average $\mathcal{S}$ of $|\psi\rangle$ is around two times greater than that of $|{\rm GHZ}_2\rangle$ \cite{Rosierpra101_2020}. For instance, $\bar{\mathcal{S}}^{2 \times 2 \times 1 \times 1}(\psi)=0.0536$ while $\bar{\mathcal{S}}^{2 \times 2}({\rm GHZ}_2)=0.028$. This observation can be easily explained based on the very definition of the inequality $I^{(N)}_{opt}$. As mentioned above, although its violation always requires $I^{(2)}_{opt}>2$, the final degree of violation depends on the $(N-2)$ single measurements. Such correction has been studied in details in Ref. \cite{Laskowskijpa48_2015}, exposing several interesting properties like higher robustness against white noise for biseparable states than for maximally entangled states. Consequently, the larger $m_3$ and $m_4$, the greater chance to find a better correction for given $m_1$ and $m_2$, what implies an increase of $\bar{\mathcal{S}}$. 

\begin{figure}
\centering
\includegraphics[width=\columnwidth]{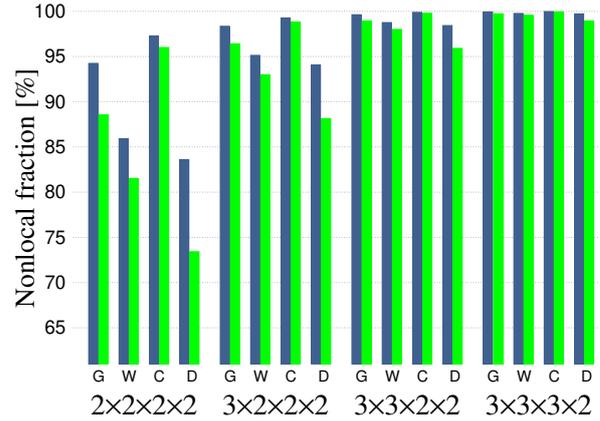}
\caption{Comparison between the nonlocal fraction calculated with linear programming method \cite{Rosierpra96_2017} (blue bars) and the nonlocal fraction estimated by mean $I^{(4)}_{opt}$ (green bars) for various measurement settings. The character G stands for $|{\rm GHZ}_4\rangle$, W for $|{\rm W}_4\rangle$, C for $|{\rm Cl}_4\rangle$, and D for $|{\rm D}^2_4\rangle$.}
\label{fig:S7a}
\end{figure}

On the other hand, for quantum states revealing multipartite entanglement, such as those in Table \ref{tab:tab4qubity}, the value of $m_3$ and $m_4$ affects both the nonlocal fraction and average nonlocal strength. Specifically, any increase in the number of measurement settings entails a fast growth of $\mathcal{P}_V$ and $\bar{\mathcal{S}}$. Consequently, for a scenario with $3 \times 3 \times 3 \times 2$ measurement settings the nonlocal fraction is close to unity. It means that our results satisfy the theorem of Lipinska et al. \cite{Lipinskanpj_2018} that $\mathcal{P}_V \to 1$ with infinitely many settings. The explanation of this fact once again has a purely statistical background, i.e. the number of settings determines the number of two-qubit state achieved after single-qubit projections and so, it becomes more likely that at least one of these states violates the CHSH inequality. Furthermore, except the simplest scenario, $2 \times 2 \times 1 \times 1$, the nonlocal fraction for multipartite entangled states surpasses $\mathcal{P}_V$ of $|\psi\rangle$. It is because, the multipartite entangled states may reveal nonlocal correlations between any pair of qubits, in contrast to the biseparable state $|\psi\rangle$.

\begin{figure}
\centering
\includegraphics[width=\columnwidth]{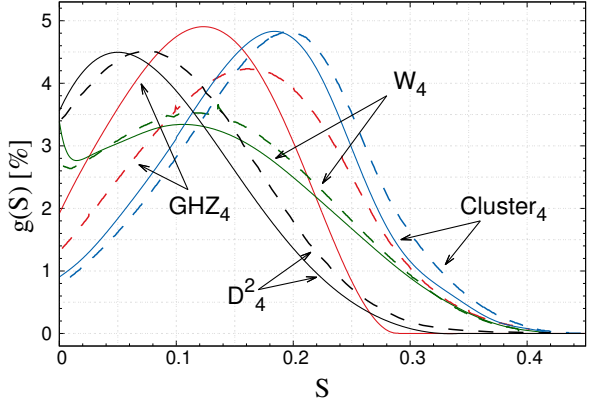}
\caption{Nonlocality strength distributions for four-qubit states with $2 \times 2 \times 2 \times 2$ measurements settings. Dashed lines denote results presented in Ref. \cite{Rosierpra101_2020}.}
\label{fig:S5}
\end{figure}

\begin{table*}
\caption{Nonlocal fraction $\mathcal{P}_V$ and average nonlocality strength $\bar{\mathcal{S}}$ for various five-qubit states and $I^{(4)}_{opt}$ inequality. }
\label{tab:tab5qubity}
%\begin{ruledtabular}
\renewcommand*{\arraystretch}{1.13}
		\begin{tabular}{@{}c@{\hspace{10pt}} c@{\hspace{10pt}} c@{\hspace{9pt}} c@{\hspace{10pt}} c@{\hspace{9pt}} c@{\hspace{10pt}} c@{\hspace{9pt}} c@{\hspace{10pt}} c@{\hspace{9pt}} c@{\hspace{10pt}} c@{\hspace{9pt}} c@{}}\toprule
 & & \multicolumn{2}{c}{${\rm GHZ}_5$} & \multicolumn{2}{c}{${\rm W}_5$} & \multicolumn{2}{c}{${\rm \operatorname{L-Cl}}_5$} & \multicolumn{2}{c}{${\rm \operatorname{R-Cl}}_5$} & \multicolumn{2}{c}{${\rm D}^2_5$}\\
\cmidrule(r){3-4}
\cmidrule(r){5-6}
\cmidrule(r){7-8}
\cmidrule(r){9-10}
\cmidrule{11-12}
Settings & Sym. & $\mathcal{P}_V[\%]$ & $\bar{\mathcal{S}}$ & $\mathcal{P}_V[\%]$ & $\bar{\mathcal{S}}$ & $\mathcal{P}_V[\%]$ & $\bar{\mathcal{S}}$ & $\mathcal{P}_V[\%]$ & $\bar{\mathcal{S}}$ & $\mathcal{P}_V[\%]$ & $\bar{\mathcal{S}}$ \\ 
\hline\hline
 $2 \times 2 \times 1 \times 1 \times 1$ & $64$ & $30.737$ & $0.0179$ & $14.932$ & $0.0126$ & $51.732$ & $0.0556$ & $41.467$ & $0.0414$ & $18.809$ & $0.0098$ \\ 
 $2 \times 2 \times 2 \times 1 \times 1$ & $384$ & $74.882$ & $0.0564$ & $51.423$ & $0.0522$ & $87.357$ & $0.1226$ & $87.424$ & $0.1209$ & $60,141$ & $0.0382$ \\
 $2 \times 2 \times 2 \times 2 \times 1$ & $1536$ & $94.347$ & $0.0998$ & $84.875$ & $0.1186$ & $97.620$ & $0.1716$ & $98.788$ & $0.1884$ & $90.637$ & $0.0809$ \\
 $2 \times 2 \times 2 \times 2 \times 2$ & $4608$ & $99.202$ & $0.1392$ & $96.661$ & $0.1848$ & $99.813$ & $0.2249$ & $99.899$ & $0.2315$ & $98.488$ & $0.1213$ \\
 \bottomrule
\end{tabular}
%\end{ruledtabular}
\end{table*}

Finally, a very interesting observation can be made when comparing our results with the previous calculations based on linear programming method \cite{Rosierpra96_2017,Rosierpra101_2020}. As illustrated in Fig. \ref{fig:S7a}, the nonlocal fraction for $I^{(4)}_{opt}$ is in quite good agreement with $\mathcal{P}_V^{{\rm L}_4}$. The best approximation of exact results is obtained for the Cluster state $|{\rm Cl}_4\rangle$, with accuracy not greater that $2~p.p.$, while the weaker estimation appears for the Dicke state $|{\rm D}^2_4\rangle$ (accuracy not greater that $10~p.p.$). Naturally, the gap between $\mathcal{P}_V^{L_4}$ and $\mathcal{P}_V^{I_{opt}}$ decreases with the number of measurement settings. It is worth mentioning that previous calculation made for the MABK and WWW\.ZB inequalities yield a much smaller results. For instance, the nonlocal fraction for $2 \times 2 \times 2 \times 2$ scenario is equal to $\mathcal{P}_V^{I_{\text{MABK}}}({\rm GHZ}_4)=13.410\%$ and $\mathcal{P}_V^{I_{\text{WWW\.ZB}}}({\rm GHZ}_4)=23.407\%$ \cite{Liangprl104_2010,Wallmanpra83_2011}.

When analyzing the histograms for the four-qubit states and $2 \times 2 \times 2 \times 2$ scenario we noticed an interesting behavior, i.e. an atypical character of the nonlocality strength distribution for the ${\rm GHZ}_4$ state (see Fig. \ref{fig:S5}). While the $g(\mathcal{S})$ functions for $|{\rm Cl}_4\rangle$, $|{\rm W}_4\rangle$, and even $|{\rm D}^2_4\rangle$ state quite well correspond to the exact results \cite{Rosierpra101_2020} (up to the slight shift towards zero), the  probability of violating $I^{(4)}_{opt}$ by the ${\rm GHZ}_4$ state strongly surpasses the exact prediction in the intermediate values of $\mathcal{S}$ and then vanishes around $\mathcal{S} = 0.29$ despite non-zero value for $g^{L_4}(\mathcal{S})$. 
This behavior is naturally reflected in the average nonlocality strength. 
For instance, for $2 \times 2 \times 2 \times 2$ scenario $\bar{\mathcal{S}}^{L_4}({\rm GHZ}_4)=0.1578$ \cite{Rosierpra101_2020} while our simulation gives $0.1067$. 
In the same time, $\bar{\mathcal{S}}^{L_4}({\rm W}_4)=0.1274$, $\bar{\mathcal{S}}^{L_4}({\rm Cl}_4)=0.1842$, and $\bar{\mathcal{S}}^{L_4}({\rm D}^2_4)=0.0954$ what agrees with our numerical results (see Table \ref{tab:tab4qubity}). 
A possible explanation of this atypical behavior could be the fact that for $|{\rm GHZ}_4\rangle$ the $I^{(4)}_{opt}$ family provides the highest violation strength in $14.6\%$ of a random set of settings while $79.04\%$ of them require a genuine $2 \times 2 \times 2 \times 2$ Bell inequality. On the contrary, for the $W_5$ state $55.06\%$ of highest violation involve $I^{(4)}_{opt}$ and just $17.68\%$ of them engage $2 \times 2 \times 2 \times 2$ measurements settings.

For the $W_5$ state it is also important to mention about a presence of a dip for nonlocality strength close to $0.02$. As we see in Fig. \ref{fig:S5}, when the analysis is restricted only to the inequality $I^{(4)}_{opt}$ the function $g(\mathcal{S})$ takes the sharper minimum compared to the entire polytope. However, due to the ambiguity of violation for $\mathcal{S} \leq 0.015$ such dip is rather meaningless from the experimental point of view.

\begin{figure}
\centering
\includegraphics[width=\columnwidth]{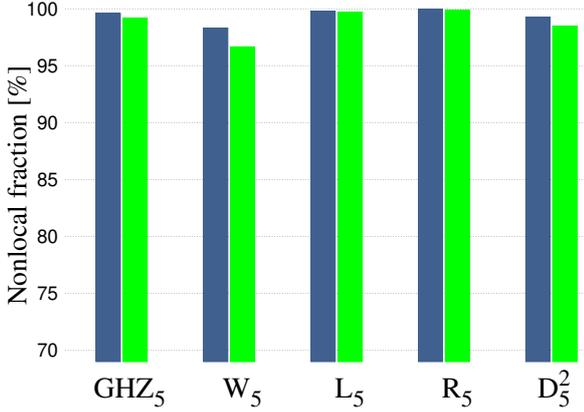}
\caption{Comparison between the nonlocal fraction calculated with linear programming method \cite{Rosierpra96_2017} (blue bars) and the nonlocal fraction estimated by mean $I^{(5)}_{opt}$ (green bars).
}
\label{fig:S8a}
\end{figure}

The results for the five-qubit states expose the above observations even more strongly. Specifically, when the number of measurement settings increases, a rapid growth of $\mathcal{P}_V$ and $\bar{\mathcal{S}}$ is observed (see Table \ref{tab:tab5qubity}). In particular, our numerical simulation shows that for a scenario with $2 \times 2 \times 2 \times 2 \times 2$ measurement settings the nonlocal fraction $\mathcal{P}_V^{I_{opt}}$ is near $100\%$ for any of the studied five-qubit states. 
Furthermore, a comparison with previous calculations based on linear programming \cite{Rosierpra96_2017} reveals a very good agreement between these two sets of outcomes, with the gap not greater then $2~p.p.$. As illustrated in Fig. \ref{fig:S8a}, the best compatibility is found for the linear- and ring-cluster states \cite{Heinpra69_2004} while the weaker estimation occurs for $|W_5\rangle$. 

\begin{figure}
\centering
\includegraphics[width=\columnwidth]{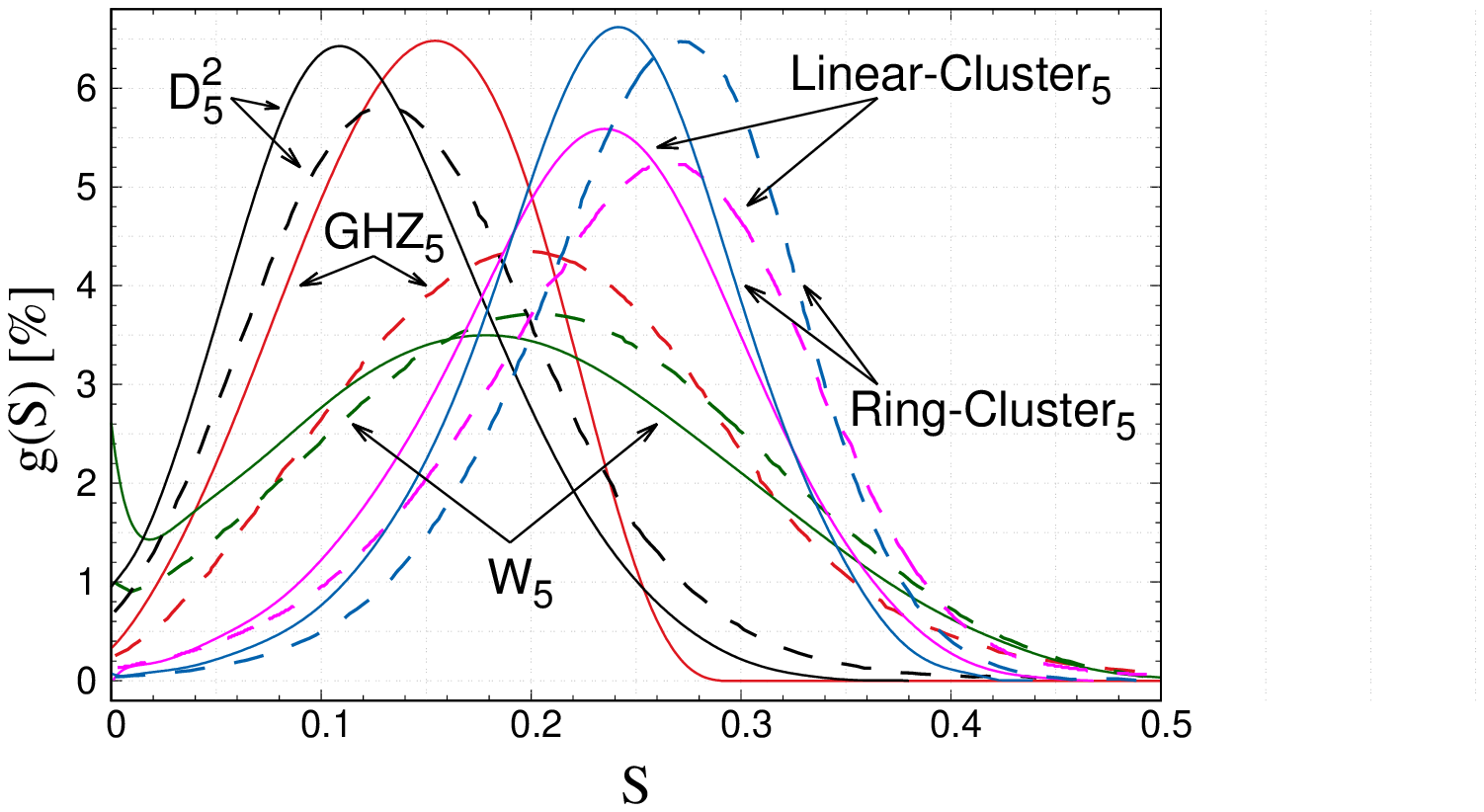}
\caption{Nonlocality strength distributions for five-qubit states with $2 \times 2 \times 2 \times 2 \times 2$ measurements settings. Dashed lines denote results presented in Ref. \cite{Rosierpra101_2020}.}
\label{fig:S6}
\end{figure}

As before, the atypical character of the nonlocality strength distribution for the GHZ state takes place also here, i.e. the functions  $g^{L_5}(\mathcal{S})$ and $g^{I_{opt}}(\mathcal{S})$ for a $2 \times 2 \times 2 \times 2 \times 2$ scenario, presented in Fig. \ref{fig:S6}, have markedly different shapes. 
For other states, $g^{I_{opt}}(\mathcal{S})$ distribution agrees qualitatively with results perform for the whole polytope, though exposing the shift towards zero (stronger than previously) and usually higher maximum compared to $g^{L_5}(\mathcal{S})$. The greater shift causes a higher difference between $\bar{\mathcal{S}}^{L_5}$ and $\bar{\mathcal{S}}^{I_{opt}}$. Specifically, 
 $\bar{\mathcal{S}}^{L_5}({\rm GHZ}_5)=0.2110$, $\bar{\mathcal{S}}^{L_5}({\rm W}_5)=0.2109$, $\bar{\mathcal{S}}^{L_5}({\rm \operatorname{L-Cl}}_5)=0.2562$,  $\bar{\mathcal{S}}^{L_5}({\rm \operatorname{R-Cl}}_5)=0.2654$, and $\bar{\mathcal{S}}^{L_5}({\rm D}^2_5)=0.1503$ what provides the difference between $\bar{\mathcal{S}}^{L_5}$ and $\bar{\mathcal{S}}^{I_{opt}}$ of around $0.072$ for the ${\rm GHZ}_5$ and less than $0.034$ for other states (see Table \ref{tab:tab5qubity}).

\subsubsection{Typicality of nonlocal correlations}

All results presented above clearly demonstrate a promising role of $I^{(N)}_{opt}$ for experimental determination of statistical aspects of nonlocal correlations. A natural question is whether for an arbitrarily generated $N-$qubit state the quality of such determination is closer to that of e.g. cluster states or rather GHZ states. Therefore, in order to make our main conclusion more general, let us study the typical nonlocal fraction $T_V$ for a randomly sampled pure state. In other words, in this problem we specify only the number of qubits $N$ and the Bell scenario, without choosing a prior any quantum state. The typical nonlocal fraction $T_V$ is given by 
\begin{equation}
T_V = \frac{1}{\Omega_{\rho}}\int \mathcal{P}_V(\rho) d\rho,
\end{equation}
where $\Omega_{\rho}$ denotes the amount of quantum states $\rho$ and the states are uniformly sampled on the surface of the Bloch sphere. By analogy, we will also compute the averaged strength $T_S$ in this more general situation.

\begin{table}
	\caption{Typical nonlocal fraction $T_V$ and typical nonlocality strength $T_S$ for pure random qubit states and random measurements. For each case (number of qubits and settings) the calculations of both quantities have been performed for $1.2*10^5$ randomly chosen states. Results for $T^{L_N}_V$ and $T^{L_N}_S$ are taken from Ref.\cite{Rosierpra101_2020}.
	}
	\label{tab:typicality}
%	\begin{ruledtabular}
\renewcommand*{\arraystretch}{1.13}
		\begin{tabular}{@{}c@{\hspace{7pt}} c@{\hspace{6pt}} c@{\hspace{7pt}} c@{\hspace{6pt}} c@{\hspace{6pt}} c@{}}\toprule
			$N$ & Settings & $T^{I^{}_{opt}}_V[\%]$ & $T^{I^{}_{opt}}_S$ & $T^{L_N}_V[\%]$ & $T^{L_N}_S$\\ 
			\hline\hline
			$3$ & $2 \times 2 \times 2$ & $38.277$ & $0.0311$ & $42.96$ & $0.034$ \\
			& $3 \times 2 \times 2$ & $57.983$ & $0.0539$ & $-$ & $-$ \\
			& $3 \times 3 \times 2$ & $76.006$ & $0.0832$ & $-$ & $-$ \\
			& $3 \times 3 \times 3$ & $88.622$ & $0.1155$ & $-$ & $-$ \\ $4$ & $2 \times 2 \times 1 \times 1$ & $19.184$ & $0.0142$ & $-$ & $-$ \\
			& $2 \times 2 \times 2 \times 1$ & $59.937$ & $0.0312$ & $-$ & $-$ \\
			& $2 \times 2 \times 2 \times 2$ & $90.096$ & $0.1118$ & $93.28$ & $0.123$ \\
			& $3 \times 2 \times 2 \times 2$ & $96.845$ & $0.1478$ & $-$ & $-$ \\
			& $3 \times 3 \times 2 \times 2$ & $99.244$ & $0.1822$ & $-$ & $-$ \\
			& $3 \times 3 \times 3 \times 2$ & $99.867$ & $0.2134$ & $-$ & $-$ \\ $5$ & $2 \times 2 \times 1 \times 1 \times 1$ & $33.801$ & $0.0264$ & $-$ & $-$ \\
			& $2 \times 2 \times 2 \times 1 \times 1$ & $80.162$ & $0.0854$ & $-$ & $-$ \\
			& $2 \times 2 \times 2 \times 2 \times 1$ & $97.327$ & $0.1512$ & $-$ & $-$ \\
			& $2 \times 2 \times 2 \times 2 \times 2$ & $99.733$ & $0.2021$ & $99.88$ & $0.222$ \\ \bottomrule
			\end{tabular} %\end{ruledtabular} 
\end{table}

As expected, the typicality $T_V$ grows as the number of settings increases, reaching the value close to $100\%$, although $m_i \leq 3$ (see Table \ref{tab:typicality}). This means that the violation of local realism can be detected for almost all states by employing Bell scenario with not more than three randomly chosen measurement settings. Furthermore, by detection we understand the violation only the very simple inequality $I^{(N)}_{opt}$. 
When comparing our results with the known outcomes for the whole polytope \cite{Rosierpra101_2020}, we see that such detection underestimate $T_V$ of around $3-4~p.p.$ and $T_S$ by about $0.02$, depending on the  number $N$. 
Naturally, when the number of observers grows the underestimation of $T_S$ further increases, after all our approach of detection is based on the strong limitation of the set of Bell inequalities. 
However, in our opinion the results presented here suffice to consider the $I^{(N)}_{opt}$ family of Bell inequalities as a simple tool for experimental studies of the subject. Furthermore, complementary to \cite{Liangprl104_2010,Wallmanpra83_2011,Rosierpra96_2017,Rosierpra101_2020}, our paper gives insight into the geometry of Bell correlations in the case of multiqubit systems, showing the the majority of the phenomenon can be explain by a scenario which effectively  involve two measurement settings.

\subsection{Multipartite Entaglement Detection}

A potential application of our findings is the detection of multipartite entanglement. As proven in \cite{Rosierpra96_2017}, if for a given state with two measurement settings per party the nonlocal fraction $\mathcal{P}_V > 2(\pi-3) \approx 28.319 \%$ then the state is multipartite entangled \cite{Rosierpra101_2020} and it cannot be written by convex combination of states which involve only two-party entanglement. Based on this fact we can conclude that some of the considered inequalities can be seen as multipartite entanglement witnesses. In the $2 \times 2 \times 2$ Bell scenario the inequalities: $I^{(3)}_{opt}, I_5, I_6, I_{13}$ and $I_{19}$ can detect genuine three qubit entanglement of $GHZ_3$ state while for the $W_3$ state the set of such inequalities is smaller and contains only the inequalities:  $I^{(3)}_{opt}, I_5$ and $I_6$. In general, the highest detection efficacy is naturally provided by the $I^{(3)}_{opt}$. For example, if one consider the state $|\psi_3\rangle$  (see Appendix~\ref{appendixB}) than $\mathcal{P}_V^{I_{opt}} = 39.970 \%$ ($\mathcal{P}_V^{L_3}(\psi_3) = 43.777\%$) whereas the second best approximation of $\mathcal{P}_V$ based on $I_6$ is equal to $22.139\%$ and so below the threshold values. 
	
When the number of measurement settings increase the threshold values of $\mathcal{P}_V$ to certify the multipartite entanglement also grows, reaching $52.401\%$ and $78.219 \%$ for $3 \times 2 \times 2$ and $3 \times 3 \times 2$ Bell scenario, respectively \cite{Rosierpra101_2020}. 
Interestingly, in these two cases the multipartite entanglement has not been detected by any Bell inequality involving genuine $m_1 \times m_2 \times m_3$ settings (c.f. Tab. \ref{tab:tab_322_results}). 
When the number of qubit increase the $I^{(N)}_{opt}$ family is sufficient to detect multipartite entanglement even for the $2 \times 2 \times 1 \times 1$ scenario. For instance, the Cluster state is detect as non two-producible (i.e. at least three partite entangled) as the nonlocal fraction are greater than the respective thresholds (see Tab. \ref{tab:entanglement}). Similar conclusions can be drawn for the five-qubit case.

\begin{table}
	\caption{
		We present the threshold values of $\mathcal{P}_V$ which can be achieved with two-producible states (i.e. states which involve only two-party entanglement). They can be calculated using the formalism presented in \cite{Rosierpra96_2017} with $\mathcal{P}^{ij}_V$ being the nonlocal fraction for $|{\rm GHZ}_2\rangle$ and the scenario $i \times j$. Other quantities are given by $\mathcal{P}^{2222}_V = 1-(1-\mathcal{P}^{22}_V)^2$, $\mathcal{P}^{3222}_V =1-(1-\mathcal{P}^{32}_V)(1-\mathcal{P}^{22}_V)$, $\mathcal{P}^{3322}_V =1-(1-\mathcal{P}^{33}_V)(1-\mathcal{P}^{22}_V)$, and $\mathcal{P}^{3332}_V = 1-(1-\mathcal{P}^{33}_V)(1-\mathcal{P}^{32}_V)$. In the last column we present examples of highly entangled states detected as non two-producible be mean of $\mathcal{P}_V$.
		}
	\label{tab:entanglement}
%	\begin{ruledtabular}
\renewcommand*{\arraystretch}{1.13}
		\begin{tabular}{@{}c@{\hspace{4pt}} c@{} c@{} c@{} c@{}} \toprule
			 &  & Max. $\mathcal{P}_V[\%]$ for & Detected &\\ 
			$N$ & Settings & two-producible states & state &\\ 
			\hline\hline
		$3$ & $2 \times 2 \times 2$ & $\mathcal{P}^{22}_V \approx 28.32$ & ${\rm GHZ}_3$,${\rm W}_3$ &\\
			& $3 \times 2 \times 2$ & $\mathcal{P}^{32}_V \approx 52.401 \%$ & \textit{u.s.} &\\
			& $3 \times 3 \times 2$ & $\mathcal{P}^{33}_V \approx 78.219$ & \textit{u.s.} &\\
			& $3 \times 3 \times 3$ & $\mathcal{P}^{33}_V$ & \textit{u.s.} &\\ 
		$4$ & $2 \times 2 \times 1 \times 1$ & $\mathcal{P}^{22}_V$ & ${\rm Cl}_4$ &\\
			& $2 \times 2 \times 2 \times 1$ & $\mathcal{P}^{22}_V$ & ${\rm Cl}_4$,${\rm GHZ}_4$,\\
			& & & ${\rm W}_4$, ${\rm D}^2_4$ &\\
			& $2 \times 2 \times 2 \times 2$ & $\mathcal{P}^{2222}_V \approx 48.62$ & \textit{u.s.} &\\
			& $3 \times 2 \times 2 \times 2$ & $\mathcal{P}^{3222}_V \approx 65.88$ & \textit{u.s.} &\\
			& $3 \times 3 \times 2 \times 2$ & $\mathcal{P}^{3322}_V \approx 84.39$ & \textit{u.s.} &\\
			& $3 \times 3 \times 3 \times 2$ & $\mathcal{P}^{3332}_V \approx 89.63$ & \textit{u.s.} &\\ 
		$5$ & $2 \times 2 \times 1 \times 1 \times 1$ & $\mathcal{P}^{22}_V$ & ${\rm GHZ}_4$, ${\rm \operatorname{L-Cl}}_5$,\\
		 & & & ${\rm \operatorname{R-Cl}}_5$ &\\
			& $2 \times 2 \times 2 \times 1 \times 1$ & $\mathcal{P}^{22}_V$ & \textit{u.s.},${\rm W}_5$,${\rm D}^2_5$ &\\
			& $2 \times 2 \times 2 \times 2 \times 1$ & $\mathcal{P}^{2222}_V$ & \textit{u.s.} &\\
			& $2 \times 2 \times 2 \times 2 \times 2$ & $\mathcal{P}^{2222}_V$ & \textit{u.s.} &\\  \bottomrule
		\end{tabular} 
%	\end{ruledtabular} 
\end{table}
		
Let us also mention the possibility of detecting genuine four-partite entanglement in these emblematic four-qubit states. For $2 \times 2 \times 2$ settings according to Tab.~\ref{tab:tab_222_results} the three-qubit GHZ state gives the threshold $\mathcal{P}_V=74.688\%$ (whereas $W$ state gives the smaller value of $\mathcal{P}_V=54.893\%$). Moreover Rosier et al.~\cite{Rosierpra96_2017} supports numerically (see Table~I in Ref.~\cite{Rosierpra96_2017}) that among all biproduct four-qubit states the largest threshold for the nonlocal fraction is $\mathcal{P}_V=74.688\%$. From this threshold it follows that any $\mathcal{P}_V$ value higher than $74.688\%$ in the $2 \times 2 \times 2 \times 2$ scenario indicates genuine four-qubit entanglement in the quantum state. Then according to Tab.~\ref{tab:tab4qubity} and relying on the validity of the this threshold, the ${\rm GHZ}_4$, ${\rm W}_4$ and the Cluster states are all detected as genuinely four-qubit entangled.
			
\section{Experimental implementation}

\subsection{Measurement device}

\begin{figure}
\centering
\includegraphics[width=\columnwidth]{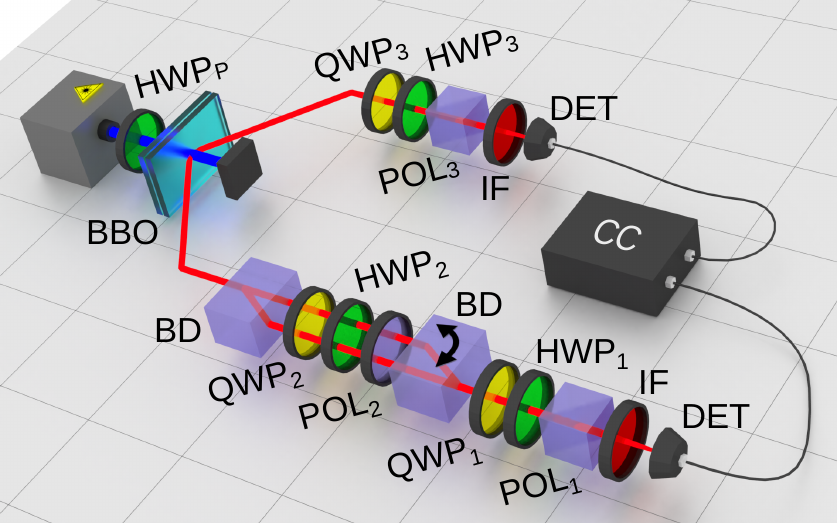}
\caption{Scheme of the experimental setup for GHZ state preparation and three-qubit projection. 
  $HWP$: half-wave plate; $QWP$: quarter-wave plate; BBO: double crystal cascade; POL: polarizer; IF: band-pass filter, BD: beam displacer, DET: detector, CC: coincidence electronics.
  \label{fig:setup}}
\end{figure}

We have experimentally tested theoretical ideas for the GHZ state contained in this paper using the platform of linear optics with discrete photons as qubit cariers (see Fig. \ref{fig:setup}). Quantum state was encoded into both their polarization state as well as into their spatial mode. To generate photons in an entangled state, we have adopted the idea by Kwiat et al. \cite{Kwiat_PRA60}. A crystal cascade is used that consists of two BBO ($\beta$-BaB$_2$O$_4$) crystals both cut for Type-I spontaneous parametric downconversion placed in contact so that their optical axes lie in perpendicular planes. The crystals are 1 mm thick and generate photon pairs at 710 nm when pumped by a laser beam at 355 nm. Coherent Paladin picosecond pulsed laser with repetition rate of 120 MHz was used in this particular experiment.

First crystal generates horizontally ($H$) polarized photons when pumped by vertically polarized laser beam. Second crystal generates vertically ($V$) polarized photon pairs when subject to horizontally polarized pumping. Rotating the laser beam polarization by an angle $\vartheta$ (with respect to the vertical polarization), we pump both the crystals coherently and obtain a superposition state
\begin{equation}
\label{eq:entangled_photons0}
%\frac{1}{\sqrt{2}} \left( 
\mathrm{cos}\vartheta |HH\rangle +
\mathrm{sin}\vartheta |VV\rangle,% \right),
\end{equation}
where letters denote polarization of the first and second photons
respectively. Note that the probability that two pairs are generated
simultaneously is negligible.

At this point we associate horizontal and vertical polarization state with logical states $|0\rangle$ and $|1\rangle$. To generate a three-qubit entangled GHZ state, we need to make use of an additional degree of freedom of the first photon and entangle this degree of freedom with the photon's polarization state. This is experimentally achieved by using a beam displacer where horizontally polarized light continues propagating along the input spatial mode, but vertically polarized light is displaced into a parallel well separated spatial mode (see beam displacer BD in Fig. \ref{fig:setup}).  
One can easily check that when labeling the original spatial mode logical $|0\rangle$ and the displaced mode $|1\rangle$, the overall state of the photon pair in (\ref{eq:entangled_photons0}) transforms into the form of
\begin{equation}
\label{eq:entangled_photons1}
%\frac{1}{\sqrt{2}} \left( 
|\vartheta\rangle = \mathrm{cos}\vartheta |000\rangle +
\mathrm{sin}\vartheta |111\rangle,% \right),
\end{equation}
where the first symbol in each bracket labels the spatial mode of the first photon, the second and third symbols correspond the polarization states of the first and second photon respectively.

To observe the correlations between individual qubits, we perform local projections on all three of them simultaneously and record coincident detections within a 2 ns interval. The polarization-encoded qubit of the first photon is projected by subjecting simultaneously both its spatial modes to a half and quarter-wave plates followed by a polarizer. Subsequently, the spatial mode encoded into the first photon is converted to its polarization mode using a second beam displacer. We recall that the original polarization qubit of the first photon has already been projected before this conversion happens. Then a sequence of half and quarter-wave plates and a polarizer is used to perform projection. The projection of the third qubit is implemented by projecting polarization state of the second photon using again, as usual,  the sequence of half and quarter-wave plates and a polarizer.

The projections have been generated independently for each qubit, together with their orthogonal counterparts.
The overall two-photon detection rate in our setup was about $64$ events per second. To have reliably enough data we accumulated the detections for $20$ second in any individual measurement setting. Including the time needed for automated measurement setting changes (wave plates rotations) acquisition of the entire data set took around $350$ hours.

\begin{table*}
\caption{Comparison between theoretical and experimental results of nonlocal fraction $\mathcal{P}_V$ and average nonlocality strength $\bar{\mathcal{S}}$. Theory refers to the states $\rho = v |\vartheta\rangle \langle \vartheta | + (1-v) \rho_{white~noise}$.
}
\label{tab:exp_results}
%\begin{ruledtabular}
\renewcommand*{\arraystretch}{1.13}
\begin{tabular}{@{}c c c c c c c c c c c@{}}\toprule
 & \multicolumn{2}{c}{$v=1$} & \multicolumn{2}{c}{$v=0.97$} & \multicolumn{2}{c}{$v=0.96$} & \multicolumn{2}{c}{$v=0.95$} & \multicolumn{2}{c}{Experiment}\\
\cmidrule(r){2-3}
\cmidrule(r){4-5}
\cmidrule(r){6-7}
\cmidrule(r){8-9}
\cmidrule{10-11}
Settings & $\mathcal{P}_V[\%]$ & $\bar{\mathcal{S}}$ & $\mathcal{P}_V[\%]$ & $\bar{\mathcal{S}}$ & $\mathcal{P}_V[\%]$ & $\bar{\mathcal{S}}$ & $\mathcal{P}_V[\%]$ & $\bar{\mathcal{S}}$ & $\mathcal{P}_V[\%]$ & $\bar{\mathcal{S}}$ \\ 
\hline\hline
 $2 \times 2 \times 2$ & $70.048$ & $0.0782$ & $61.183$ & $0.0603$ & $57.990$ & $0.0548$ & $54.863$ & $0.0494$ & $56 \pm 5$ & $0.059 \pm 0.001$ \\ 
 $3 \times 2 \times 2$ & $87.512$ & $0.0962$ & $81.343$ & $0.0955$ & $78.880$ & $0.0877$ & $76.073$ & $0.0806$ & $75 \pm 4$ & $0.0897 \pm 0.0007$ \\
 $3 \times 3 \times 2$ & $96.173$ & $0.1413$ & $93.424$ & $0.1321$ & $92.110$ & $0.1239$ & $90.641$ & $0.1151$ & $88 \pm 2$ & $0.1234 \pm 0.0005$ \\
 $3 \times 3 \times 3$ & $99.268$ & $0.1846$ & $98.440$ & $0.1659$ & $98.032$ & $0.1571$ & $97.569$ & $0.1487$ & $95 \pm 2$ & $0.1553 \pm 0.0002$ \\ \bottomrule
\end{tabular}
%\end{ruledtabular}
\end{table*}

\subsection{Experimental results for random measurements}

The experiment is composed of two steps. In the first one, we use the setup presented in Fig. \ref{fig:setup} to
prepare the three-qubit GHZ states. It is know that in any experimental preparation of the quantum state, various kinds of imperfections are inevitably present. The imperfections are caused, e.g., by improper setting of individual experimental components or by depolarization effects (presence of noise). To get an information about their presence in the generated state $\rho_{expt}$, the quantum state tomography and maximum-likelihood estimation have been used to reconstruct the output-state density matrix \cite{Jezek_PRA68, PRL100}. From this data the output-state can be approximated by
\begin{equation}
\rho_{expt} = v |\vartheta\rangle \langle \vartheta | + (1-v)\rho_{\rm white~noise},
\end{equation}
where the parameter $\vartheta = 45^{\circ}$ with a precision of $\pm0.5^{\circ}$ and the  visibility $v = 0.97 \pm 0.01$. Notice, that this approximation models all experimental imperfections in terms of white noise.  
The error bars are determined by Monte Carlo simulations of Poissonian noise distribution. Note that $\rho_{expt}$ is a perfect theoretical density matrix of a Werner state with only the value of $v$ obtained from experimental tomography.

Next, for the state $\rho_{expt}$ the violation of $I^{(3)}_{opt}$ family has been studied for $n=5 \cdot 10^4$ randomly generated sets of measurement settings. This quantity ensures sufficiently dense sampling. Each set denotes an ensemble of projective measurement $E^i_{k_i}=\vec{e}^i_{k_i} \cdot \vec{\sigma}$, where $i=1,2,3$, $\vec{\sigma}=\{\sigma_x,\sigma_y,\sigma_z\}$ corresponds to the vector of the Pauli operators associated with three orthogonal directions, and all unit vectors are represent in spherical coordinates, $\vec{e}^i_{k_i}=(\sin 2\phi^i_{k_i} \cos \xi^i_{k_i}, \sin 2\phi^i_{k_i} \sin \xi^i_{k_i}, \cos 2\phi^i_{k_i})$. The projective measurement are generated by random sampling the angles $\{\xi^i_{k_i}, \phi^i_{k_i}\}$ according to the Haar measure \cite{Zyczkowskijpa27_1994}, namely, $\xi^i_{k_i}$ is taken from uniform distribution on the intervals $\langle 0, 2 \pi)$, while $\phi^i_{k_i}= \arcsin(\sqrt{\omega^i_{k_i}})$ and $\omega^i_{k_i}$ is distributed uniformly on $\langle 0, 1)$. All variables are generated independently for each observer and  measurement number $k_i$. 
To measure any correlation coefficient $\langle E^{1}_{k_1} \cdots E^{N}_{k_N}\rangle$ the six wave plates in the projection part of the setup are adjusted accordingly to the angles $\{\xi^i_{k_i}, \phi^i_{k_i}\}$. 
For each sets of angles, the coincidence counts on the two detectors are measured for a fixed amount of time and then, the maximal value of $I^{(3)}_{opt}$ family is computed, taking into account all permutation of parties, inputs and outputs as detailed above. The value of $I^{(3)}_{opt}$ is determined with precision $\pm 0.015$. Dividing the number of detected violation of local realism by $n$, the nonlocal fraction is estimated. Similarly, the average nonlocality strength is estimated.

\begin{figure}
\centering
\includegraphics[width=\columnwidth]{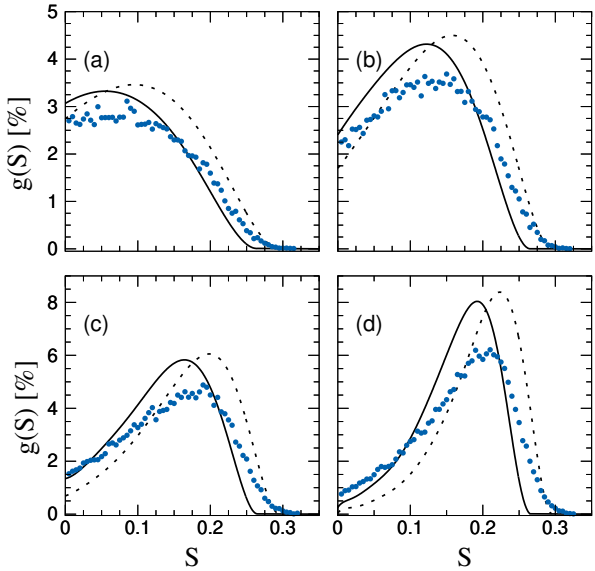}
\caption{Comparison between theoretical and experimental estimation of the nonlocality strength distribution for various scenario: (a) $2 \times 2 \times 2$ measurement settings; (b) $3 \times 2 \times 2$ measurement settings;
(c) $3 \times 3 \times 2$ measurement settings;
(d) $3 \times 3 \times 3$ measurement settings. 
Dashed lines denote theoretical prediction for pure three-qubit GHZ state while the solid lines corresponds to $\rho = 0.96 |\vartheta\rangle \langle \vartheta | + 0.04 \rho_{white~noise}$. Circle symbols depict experimental results.}
\label{fig:Exp1}
\end{figure}

The experimental results are collected in Table~\ref{tab:exp_results}. As we see, our measurements are in good agreement with theoretical predictions for $v=0.96$. It proves strong nonlocal properties of the GHZ state. Since the nonlocal fraction for the scenario $2 \times 2 \times 2$, $\mathcal{P}_V = 56\pm 5$, is clearly greater than $2(\pi-3)\approx 28.319$ the experiment revealed genuine three gubit entanglement of the GHZ state. Furthermore, to present more details about our experimental results, the histograms of the nonlocality strength have been studied. The experimental distributions are presented in Fig. \ref{fig:Exp1}, where each point denotes $g^{expt}(\mathcal{S})$ for the nonlocality strength in the interval $(\mathcal{S},\mathcal{S}+0.005)$. For all analyzed $m_1\times m_2 \times m_3$ scenario, the function $g^{expt}(\mathcal{S})$ has a similar shape as its theoretical counterpart, despite reaching the smallest maximal value. In general, the higher the value of $S$, the better the experimental data fits its theoretical prediction since data with higher nonlocality strength seems to be more reliable. This effect could be explained by the fact that white noise is only a first approximation of the imperfections occurring in the experimental setup. Furthermore, the wave plates (6 in total) are subject to experimental imperfections, especially measurable when using simultaneously all of them at completely random settings. This explains the slight discrepancy between theoretical predication and the experiment when measuring the g(S) quantity. Note that quantum tomography seams more robust again this sort of problems because of the limited set of wave plates settings used to obtain it.

\section{Conclusions}

In this paper we investigated the nonlocal fraction and the nonlocality strength as two important quantities characterizing nonlocal correlations of the quantum states. 
The overall message of the obtained results is that both quantities can be accurately estimated using a greatly simplified model of nonlocality based only on the violation of one class of lifted CHSH inequality, namely $I^{(N)}_{opt}$. A simple form of $I^{(N)}_{opt}$ expressed by the correlation coefficients makes these inequalities of paramount importance for practical experimental investigation of all problems discussed in this paper. In particular, the nonlocal fraction can be used as a witness of genuine multipartite entanglement without having the distant parties share a common reference frames. In contrast to the witnesses based on the MABK and WWW\.ZB inequalities, our procedure provides a significant increase in entanglement-detection efficiency.

On the other hand, our results shed a new light on the geometry of the quantum correlations, showing that statistically the most relevant one involve effectively two measurements settings per party. In other words, the lifted CHSH inequality is the first approximation of the quantum correlations polytope. The statistical relevance of the lifted CHSH inequality has been numerically shown for various number of measurement setting. This result implies that the increase of the nonlocal fraction towards unity when the number of measurement choices tends to infinity originates in statistical reasons i.e. a greater number of equivalent the lifted CHSH inequalities.

\section{Acknowledgements}
The authors thank Cesnet for providing data management services. Numerical calculations were performed in the Wroclaw Centre for Networking and Supercomputing, Poland. A.B., K.L., and J.S. are supported by GA \v{C}R Project No. 20-17765S. A.B. also thanks M\v{S}MT \v{C}R for support by the Project No. CZ.02.1.01/0.0/0.0/16$\_$019/0000754 and No. CZ.1.05/2.1.00/19.0377. WL acknowledges support from the Foundation for Polish Science (IRAP project ICTQT, Contract No. 2018/MAB/5, cofinanced by EU via Smart Growth Operational Programme). TV acknowledges the support of the EU (QuantERA eDICT) and the National Research, Development and Innovation Office NKFIH (No. 2019-2.1.7-ERA-NET-2020-00003).

%\bibliography{citace}

\begin{thebibliography}{47}
\providecommand{\natexlab}[1]{#1}
\providecommand{\url}[1]{\texttt{#1}}
\expandafter\ifx\csname urlstyle\endcsname\relax
  \providecommand{\doi}[1]{doi: #1}\else
  \providecommand{\doi}{doi: \begingroup \urlstyle{rm}\Url}\fi

\bibitem[Bell(1964)]{Bellphysics1_1964}
J.~S. Bell.
\newblock On the eistein podolsky rosen paradox.
\newblock \emph{Physics}, 1:\penalty0 195, 1964.
\newblock \doi{10.1103/PhysicsPhysiqueFizika.1.195}.

\bibitem[Einstein et~al.(1935)Einstein, Podolsky, and Rosen]{Einsteinpr47_1935}
A.~Einstein, B.~Podolsky, and N.~Rosen.
\newblock Can quantum-mechanical description of physical reality be considered
  complete?
\newblock \emph{Phys. Rev.}, 47:\penalty0 777, 1935.
\newblock \doi{10.1103/PhysRev.47.777}.

\bibitem[Aspect(1999)]{Aspectnature398_1999}
A.~Aspect.
\newblock Bell's inequality test: more ideal than ever.
\newblock \emph{Nature}, 398:\penalty0 189, 1999.
\newblock \doi{https://doi.org/10.1038/18296}.

\bibitem[Brunner et~al.(2014)Brunner, Cavalcanti, Pironio, Scarani, and
  Wehner]{Brunnerrmp86_2014}
N.~Brunner, D.~Cavalcanti, S.~Pironio, V.~Scarani, and S.~Wehner.
\newblock Bell nonlocality.
\newblock \emph{Rev. Mod. Phys.}, 86:\penalty0 419, 2014.
\newblock \doi{10.1103/RevModPhys.86.419}.

\bibitem[Bennet and Brassard(1984)]{Bennet_proc}
C.~H. Bennet and G.~Brassard.
\newblock Quantum cryptography: Public key distribution and coin tossing.
\newblock In \emph{Proceedings of IEEE International Conference on Computers,
  Systems and Signal Processing}, page 175, New York, USA, 1984. IEEE.

\bibitem[Ekert(1991)]{Ekertprl67_1991}
Artur~K. Ekert.
\newblock Quantum cryptography based on bell's theorem.
\newblock \emph{Phys. Rev. Lett.}, 67:\penalty0 661, 1991.
\newblock \doi{10.1103/PhysRevLett.67.661}.

\bibitem[Buhrman et~al.(2010)Buhrman, Cleve, Massar, and
  de~Wolf]{Buhrmanrmp82_2010}
Harry Buhrman, Richard Cleve, Serge Massar, and Ronald de~Wolf.
\newblock Nonlocality and communication complexity.
\newblock \emph{Rev. Mod. Phys.}, 82:\penalty0 665, 2010.
\newblock \doi{10.1103/RevModPhys.82.665}.

\bibitem[Pironio et~al.(2010)Pironio, Acín, Massar, de~la Giroday,
  Matsukevich, Maunz, Olmschenk, Hayes, Luo, Manning, and
  Monroe]{Pironionature464_2010}
S.~Pironio, A.~Acín, S.~Massar, A.~Boyer de~la Giroday, D.~N. Matsukevich,
  P.~Maunz, S.~Olmschenk, D.~Hayes, L.~Luo, T.~A. Manning, and C.~Monroe.
\newblock Random numbers certified by bell’s theorem.
\newblock \emph{Nature}, 464:\penalty0 1021, 2010.
\newblock \doi{https://doi.org/10.1038/nature09008}.

\bibitem[Ac\'{\i}n et~al.(2007)Ac\'{\i}n, Brunner, Gisin, Massar, Pironio, and
  Scarani]{Acinprl98_2007}
A.~Ac\'{\i}n, N.~Brunner, N.~Gisin, S.~Massar, S.~Pironio, and V.~Scarani.
\newblock Device-independent security of quantum cryptography against
  collective attacks.
\newblock \emph{Phys. Rev. Lett.}, 98:\penalty0 230501, 2007.
\newblock \doi{10.1103/PhysRevLett.98.230501}.

\bibitem[Rabelo et~al.(2011)Rabelo, Ho, Cavalcanti, Brunner, and
  Scarani]{Rabeloprl107_2011}
Rafael Rabelo, Melvyn Ho, Daniel Cavalcanti, Nicolas Brunner, and Valerio
  Scarani.
\newblock Device-independent certification of entangled measurements.
\newblock \emph{Phys. Rev. Lett.}, 107:\penalty0 050502, 2011.
\newblock \doi{10.1103/PhysRevLett.107.050502}.

\bibitem[M\'{e}thot and Scarani(2007)]{Methotqic7_2007}
A.~A. M\'{e}thot and V.~Scarani.
\newblock An anomaly of non-locality.
\newblock \emph{Quantum Inf. Comput.}, 7:\penalty0 157, 2007.

\bibitem[Liang et~al.(2010)Liang, Harrigan, Bartlett, and
  Rudolph]{Liangprl104_2010}
Y.-C. Liang, N.~Harrigan, S.~D. Bartlett, and T.~Rudolph.
\newblock Nonclassical correlations from randomly chosen local measurements.
\newblock \emph{Phys. Rev. Lett.}, 104:\penalty0 050401, 2010.
\newblock \doi{10.1103/PhysRevLett.104.050401}.

\bibitem[Wallman et~al.(2011)Wallman, Liang, and Bartlett]{Wallmanpra83_2011}
J.~J. Wallman, Y.-C. Liang, and S.~D. Bartlett.
\newblock Generating nonclassical correlations without fully aligning
  measurements.
\newblock \emph{Phys. Rev. A}, 83:\penalty0 022110, 2011.
\newblock \doi{10.1103/PhysRevA.83.022110}.

\bibitem[Fonseca and Parisio(2015)]{Fonsecapra92_2015}
E.~A. Fonseca and F.~Parisio.
\newblock Measure of nonlocality which is maximal for maximally entangled
  qutrits.
\newblock \emph{Phys. Rev. A}, 92:\penalty0 030101(R), 2015.
\newblock \doi{10.1103/PhysRevA.92.030101}.

\bibitem[de~Rosier et~al.(2017)de~Rosier, Gruca, Parisio, V\'ertesi, and
  Laskowski]{Rosierpra96_2017}
A.~de~Rosier, J.~Gruca, F.~Parisio, T.~V\'ertesi, and W.~Laskowski.
\newblock Multipartite nonlocality and random measurements.
\newblock \emph{Phys. Rev. A}, 96:\penalty0 012101, 2017.
\newblock \doi{10.1103/PhysRevA.96.012101}.

\bibitem[Barasi\ifmmode~\acute{n}\else \'{n}\fi{}ski and
  Nowotarski(2018)]{Barasinskipra98_2018}
Artur Barasi\ifmmode~\acute{n}\else \'{n}\fi{}ski and Mateusz Nowotarski.
\newblock Volume of violation of bell-type inequalities as a measure of
  nonlocality.
\newblock \emph{Phys. Rev. A}, 98:\penalty0 022132, 2018.
\newblock \doi{10.1103/PhysRevA.98.022132}.

\bibitem[Fonseca et~al.(2018)Fonseca, de~Rosier, V\'ertesi, Laskowski, and
  Parisio]{Fonsecapra98_2018}
Alejandro Fonseca, Anna de~Rosier, Tam\'as V\'ertesi, Wies\l{}aw Laskowski, and
  Fernando Parisio.
\newblock Survey on the bell nonlocality of a pair of entangled qudits.
\newblock \emph{Phys. Rev. A}, 98:\penalty0 042105, 2018.
\newblock \doi{10.1103/PhysRevA.98.042105}.

\bibitem[Barasi\ifmmode~\acute{n}\else \'{n}\fi{}ski
  et~al.(2020)Barasi\ifmmode~\acute{n}\else \'{n}\fi{}ski,
  \ifmmode~\check{C}\else \v{C}\fi{}ernoch, Lemr, and
  Soubusta]{Barasinskipra101_2020}
Artur Barasi\ifmmode~\acute{n}\else \'{n}\fi{}ski, Anton\'{\i}n
  \ifmmode~\check{C}\else \v{C}\fi{}ernoch, Karel Lemr, and Jan Soubusta.
\newblock Genuine tripartite nonlocality for random measurements in
  greenberger-horne-zeilinger-class states and its experimental test.
\newblock \emph{Phys. Rev. A}, 101:\penalty0 052109, 2020.
\newblock \doi{10.1103/PhysRevA.101.052109}.

\bibitem[Lipinska et~al.(2018)Lipinska, Curchod, M\'{a}ttar, and
  Ac\'{i}n]{Lipinskanpj_2018}
V.~Lipinska, F.~Curchod, A.~M\'{a}ttar, and A.~Ac\'{i}n.
\newblock Towards an equivalence between maximal entanglement and maximal
  quantum nonlocality.
\newblock \emph{New J. Phys.}, 20:\penalty0 063043, 2018.
\newblock \doi{10.1088/1367-2630/aaca22}.

\bibitem[de~Rosier et~al.(2020)de~Rosier, Gruca, Parisio, V\'ertesi, and
  Laskowski]{Rosierpra101_2020}
Anna de~Rosier, Jacek Gruca, Fernando Parisio, Tam\'as V\'ertesi, and
  Wies\l{}aw Laskowski.
\newblock Strength and typicality of nonlocality in multisetting and
  multipartite bell scenarios.
\newblock \emph{Phys. Rev. A}, 101:\penalty0 012116, 2020.
\newblock \doi{10.1103/PhysRevA.101.012116}.

\bibitem[Cope and Colbeck(2019)]{Copepra100_2019}
Thomas Cope and Roger Colbeck.
\newblock Bell inequalities from no-signaling distributions.
\newblock \emph{Phys. Rev. A}, 100:\penalty0 022114, 2019.
\newblock \doi{10.1103/PhysRevA.100.022114}.

\bibitem[Yang et~al.(2020)Yang, Tabia, Lin, and Liang]{Yangpra102_2020}
Shih-Xian Yang, Gelo~Noel Tabia, Pei-Sheng Lin, and Yeong-Cherng Liang.
\newblock Device-independent certification of multipartite entanglement using
  measurements performed in randomly chosen triads.
\newblock \emph{Phys. Rev. A}, 102:\penalty0 022419, 2020.
\newblock \doi{10.1103/PhysRevA.102.022419}.

\bibitem[Gruca et~al.(2010)Gruca, Laskowski, \ifmmode~\dot{Z}\else
  \.{Z}\fi{}ukowski, Kiesel, Wieczorek, Schmid, and
  Weinfurter]{Grucapra82_2010}
Jacek Gruca, Wies\l{}aw Laskowski, Marek \ifmmode~\dot{Z}\else
  \.{Z}\fi{}ukowski, Nikolai Kiesel, Witlef Wieczorek, Christian Schmid, and
  Harald Weinfurter.
\newblock Nonclassicality thresholds for multiqubit states: Numerical analysis.
\newblock \emph{Phys. Rev. A}, 82:\penalty0 012118, 2010.
\newblock \doi{10.1103/PhysRevA.82.012118}.

\bibitem[Hradil(1997)]{Hradilpra55_1997}
Z.~Hradil.
\newblock Quantum-state estimation.
\newblock \emph{Phys. Rev. A}, 55:\penalty0 R1561, 1997.
\newblock \doi{10.1103/PhysRevA.55.R1561}.

\bibitem[Lin et~al.(2018)Lin, Rosset, Zhang, Bancal, and Liang]{Linpra97_2018}
Pei-Sheng Lin, Denis Rosset, Yanbao Zhang, Jean-Daniel Bancal, and Yeong-Cherng
  Liang.
\newblock Device-independent point estimation from finite data and its
  application to device-independent property estimation.
\newblock \emph{Phys. Rev. A}, 97:\penalty0 032309, 2018.
\newblock \doi{10.1103/PhysRevA.97.032309}.

\bibitem[Shadbolt et~al.(2012)Shadbolt, V\'ertesi, Liang, Branciard, Brunner,
  and O’Brien]{ShadboltSciRep2_2012}
P.~Shadbolt, T.~V\'ertesi, Y.-C. Liang, C.~Branciard, N.~Brunner, and J.~L.
  O’Brien.
\newblock Guaranteed violation of a bell inequality without aligned reference
  frames or calibrated devices.
\newblock \emph{Sci. Rep.}, 2:\penalty0 470, 2012.
\newblock \doi{https://doi.org/10.1038/srep00470}.

\bibitem[Slater et~al.(2014)Slater, Branciard, Brunner, and
  Tittel]{Slaternjp16_2014}
Joshua~A Slater, Cyril Branciard, Nicolas Brunner, and Wolfgang Tittel.
\newblock Device-dependent and device-independent quantum key distribution
  without a shared reference frame.
\newblock \emph{New J. Phys.}, 16:\penalty0 043002, 2014.
\newblock \doi{10.1088/1367-2630/16/4/043002}.

\bibitem[Clauser et~al.(1969)Clauser, Horne, Shimony, and Holt]{CHSHprl23_1969}
J.~F. Clauser, M.~A. Horne, A.~Shimony, and R.~A. Holt.
\newblock Proposed experiment to test local hidden-variable theories.
\newblock \emph{Phys. Rev. Lett.}, 23:\penalty0 880, 1969.
\newblock \doi{10.1103/PhysRevLett.23.880}.

\bibitem[Pitowsky and Svozil(2001)]{Pitowskypra64_2001}
I.~Pitowsky and K.~Svozil.
\newblock Optimal tests of quantum nonlocality.
\newblock \emph{Phys. Rev. A}, 64:\penalty0 014102, 2001.
\newblock \doi{10.1103/PhysRevA.64.014102}.

\bibitem[\'Sliwa(2003)]{Sliwapla317_2003}
C.~\'Sliwa.
\newblock Symmetries of the bell correlation inequalities.
\newblock \emph{Phys. Lett. A}, 317:\penalty0 165, 2003.
\newblock \doi{10.1016/S0375-9601(03)01115-0}.

\bibitem[Boyd and Vandenberghe(2004)]{boyd_vandenberghe_2004}
Stephen Boyd and Lieven Vandenberghe.
\newblock \emph{Convex Optimization}.
\newblock Cambridge University Press, 2004.

\bibitem[Pitowsky(1989)]{PitowskyLNP_1989}
I.~Pitowsky.
\newblock \emph{Quantum Probability — Quantum Logic}, volume 321.
\newblock Springer-Verlag Berlin Heidelberg, 1989.

\bibitem[Pironio(2005)]{Pironiojmp46_2005}
S.~Pironio.
\newblock Lifting bell inequalities.
\newblock \emph{J. Math. Phys.}, 46:\penalty0 062112, 2005.
\newblock \doi{https://doi.org/10.1063/1.1928727}.

\bibitem[Laskowski et~al.(2015)Laskowski, V\'ertesi, and
  Wie\'sniak]{Laskowskijpa48_2015}
W.~Laskowski, T.~V\'ertesi, and M.~Wie\'sniak.
\newblock Highly noise resistant multiqubit quantum correlations.
\newblock \emph{J. Phys. A: Math. Theor.}, 48:\penalty0 465301, 2015.
\newblock \doi{https://doi.org/10.1088/1751-8113/48/46/465301}.

\bibitem[Kostrzewa et~al.(2018)Kostrzewa, Laskowski, and
  V\'ertesi]{Kostrzewapra98_2018}
Kamil Kostrzewa, Wies\l{}aw Laskowski, and Tam\'as V\'ertesi.
\newblock Closing the detection loophole in multipartite bell experiments with
  a limited number of efficient detectors.
\newblock \emph{Phys. Rev. A}, 98:\penalty0 012138, 2018.
\newblock \doi{10.1103/PhysRevA.98.012138}.

\bibitem[Curchod et~al.(2015)Curchod, Gisin, and Liang]{Curchodpra91_2015}
F.~J. Curchod, N.~Gisin, and Y.-C. Liang.
\newblock Quantifying multipartite nonlocality via the size of the resource.
\newblock \emph{Phys. Rev. A}, 91:\penalty0 012121, 2015.
\newblock \doi{10.1103/PhysRevA.91.012121}.

\bibitem[Jebarathinam et~al.(2019)Jebarathinam, Hung, Chen, and
  Liang]{Jebarathinamprr1_2019}
C.~Jebarathinam, Jui-Chen Hung, Shin-Liang Chen, and Yeong-Cherng Liang.
\newblock Maximal violation of a broad class of bell inequalities and its
  implication on self-testing.
\newblock \emph{Phys. Rev. Research}, 1:\penalty0 033073, 2019.
\newblock \doi{10.1103/PhysRevResearch.1.033073}.

\bibitem[Matou\v{s}ek et~al.(1996)Matou\v{s}ek, Sharir, and
  Welzl]{Matousek16_1996}
J.~Matou\v{s}ek, M.~Sharir, and E.~Welzl.
\newblock A subexponential bound for linear programming.
\newblock \emph{Algorithmica}, 16:\penalty0 498, 1996.
\newblock \doi{https://doi.org/10.1007/BF01940877}.

\bibitem[D\"ur et~al.(2000)D\"ur, Vidal, and Cirac]{Durpra62_2000}
W.~D\"ur, G.~Vidal, and J.~I. Cirac.
\newblock Three qubits can be entangled in two inequivalent ways.
\newblock \emph{Phys. Rev. A}, 62:\penalty0 062314, 2000.
\newblock \doi{10.1103/PhysRevA.62.062314}.

\bibitem[Barasi\'nski et~al.(2019)Barasi\'nski, \v{C}ernoch, Lemr, and
  Soubusta]{Barasinskipra99_2019}
A.~Barasi\'nski, A.~\v{C}ernoch, K.~Lemr, and J.~Soubusta.
\newblock Experimental verification of time-order-dependent correlations in
  three-qubit greenberger-horne-zeilinger-class states.
\newblock \emph{Phys. Rev. A}, 99:\penalty0 042123, 2019.
\newblock \doi{10.1103/PhysRevA.99.042123}.

\bibitem[Werner and Wolf(2001)]{Wernerpra64_2001}
R.~F. Werner and M.~M. Wolf.
\newblock All-multipartite bell-correlation inequalities for two dichotomic
  observables per site.
\newblock \emph{Phys. Rev. A}, 64:\penalty0 032112, 2001.
\newblock \doi{10.1103/PhysRevA.64.032112}.

\bibitem[Collins and Gisin(2004)]{Collins_2004}
D.~Collins and N.~Gisin.
\newblock A relevant two qubit bell inequality inequivalent to the {CHSH}
  inequality.
\newblock \emph{J. Phys. A}, 37:\penalty0 1775, 2004.
\newblock \doi{https://doi.org/10.1088/0305-4470/37/5/021}.

\bibitem[Hein et~al.(2004)Hein, Eisert, and Briegel]{Heinpra69_2004}
M.~Hein, J.~Eisert, and H.~J. Briegel.
\newblock Multiparty entanglement in graph states.
\newblock \emph{Phys. Rev. A}, 69:\penalty0 062311, 2004.
\newblock \doi{10.1103/PhysRevA.69.062311}.

\bibitem[Kwiat et~al.(1999)Kwiat, Waks, White, Appelbaum, and
  Eberhard]{Kwiat_PRA60}
Paul~G. Kwiat, Edo Waks, Andrew~G. White, Ian Appelbaum, and Philippe~H.
  Eberhard.
\newblock Ultrabright source of polarization-entangled photons.
\newblock \emph{Phys. Rev. A}, 60:\penalty0 R773, 1999.
\newblock \doi{10.1103/PhysRevA.69.062311}.

\bibitem[Je\ifmmode~\check{z}\else \v{z}\fi{}ek
  et~al.(2003)Je\ifmmode~\check{z}\else \v{z}\fi{}ek,
  Fiur\'a\ifmmode~\check{s}\else \v{s}\fi{}ek, and Hradil]{Jezek_PRA68}
Miroslav Je\ifmmode~\check{z}\else \v{z}\fi{}ek, Jarom\'{\i}r
  Fiur\'a\ifmmode~\check{s}\else \v{s}\fi{}ek, and Zden\ifmmode
  \check{e}\else~\v{e}\fi{}k Hradil.
\newblock Quantum inference of states and processes.
\newblock \emph{Phys. Rev. A}, 68:\penalty0 012305, 2003.
\newblock \doi{10.1103/PhysRevA.68.012305}.

\bibitem[\ifmmode~\check{C}\else \v{C}\fi{}ernoch
  et~al.(2008)\ifmmode~\check{C}\else \v{C}\fi{}ernoch, Soubusta, Bart\ifmmode
  \mathring{u}\else \r{u}\fi{}\ifmmode~\check{s}\else \v{s}\fi{}kov\'a,
  Du\ifmmode~\check{s}\else \v{s}\fi{}ek, and Fiur\'a\ifmmode~\check{s}\else
  \v{s}\fi{}ek]{PRL100}
Anton\'{\i}n \ifmmode~\check{C}\else \v{C}\fi{}ernoch, Jan Soubusta, Lucie
  Bart\ifmmode \mathring{u}\else \r{u}\fi{}\ifmmode~\check{s}\else
  \v{s}\fi{}kov\'a, Miloslav Du\ifmmode~\check{s}\else \v{s}\fi{}ek, and
  Jarom\'{\i}r Fiur\'a\ifmmode~\check{s}\else \v{s}\fi{}ek.
\newblock Experimental realization of linear-optical partial swap gates.
\newblock \emph{Phys. Rev. Lett.}, 100:\penalty0 180501, 2008.
\newblock \doi{10.1103/PhysRevLett.100.180501}.

\bibitem[\.Zyczkowski and Ku\'s(1994)]{Zyczkowskijpa27_1994}
K.~\.Zyczkowski and M.~Ku\'s.
\newblock Random unitary matrices.
\newblock \emph{J. Phys. A}, 27:\penalty0 4235, 1994.
\newblock \doi{https://doi.org/10.1088/0305-4470/27/12/028}.

\end{thebibliography}

\onecolumngrid
\newpage
\appendix

\setcounter{table}{0}
\renewcommand{\thetable}{B.\arabic{table}}
 
\section{States under considerations}
\label{appendixB}
Below we present the set of states for which statistical properties of the nonlocality strength have been analyzed
\begin{eqnarray}
|{\rm GHZ}_2 \rangle &=& (|00\rangle + |11\rangle)/\sqrt{2}, \nonumber\\
|{\rm GHZ}_3 \rangle &=& (|000\rangle + |111\rangle)/\sqrt{2}, \nonumber\\
|{\rm W}_3 \rangle &=& (|001\rangle + |010\rangle + |100\rangle)/\sqrt{3}, \nonumber\\
|{\rm GHZ}_4 \rangle &=& (|0000\rangle + |1111\rangle)/\sqrt{2}, \nonumber\\
|{\rm W}_4 \rangle &=& (|0001\rangle + |0010\rangle + |0100\rangle + |1000\rangle)/2, \nonumber\\
|{\rm D}^2_4 \rangle &=& (|0011\rangle + |0101\rangle + |0110\rangle + |1001\rangle + |1010\rangle + |1100\rangle)/\sqrt{6},\nonumber\\
|{\rm Cl}_4 \rangle &=& (|0000\rangle + |1100\rangle + |0011\rangle - |1111\rangle)/2,\nonumber\\
|{\rm GHZ}_5 \rangle &=& (|00000\rangle + |11111\rangle)/\sqrt{2},\nonumber\\
|{\rm W}_5 \rangle &=& (|00001\rangle + |00010\rangle + |00100\rangle + |01000\rangle + |10000\rangle)/\sqrt{5},\nonumber\\
|{\rm D}^2_5 \rangle &=& (|00011\rangle + |00101\rangle + |00110\rangle + |01001\rangle + |01010\rangle + |01100\rangle + |10001\rangle + |10010\rangle \nonumber\\
&&+ |10100\rangle + |11000\rangle)/\sqrt{10},\nonumber\\
|{\rm \operatorname{L-Cl}}_5 \rangle &=& (|00000\rangle + |00010\rangle + |00101\rangle - |00111\rangle + |01000\rangle + |01010\rangle + |01101\rangle - |01111\rangle \nonumber\\
&&+ |10001\rangle - |10011\rangle + |10100\rangle + |10110\rangle - |11001\rangle + |11011\rangle - |11100\rangle - |11110\rangle)/4,\nonumber\\
|{\rm \operatorname{R-Cl}}_5 \rangle &=& (|00001\rangle + |00010\rangle + |00100\rangle - |00111\rangle + |01000\rangle + |01011\rangle + |01101\rangle - |01110\rangle \nonumber\\
&& + |10000\rangle - |10011\rangle + |10101\rangle + |10110\rangle - |11001\rangle + |11010\rangle - |11100\rangle - |11111\rangle)/4,\nonumber\\
|\psi_3\rangle &=& 0.522 |000\rangle + 0.692 e^{2.387 i} |001\rangle +0.172 e^{-2.972 i} |010\rangle +0.140 e^{-0.102 i} |011\rangle \nonumber\\
&&  +0.296 e^{2.864 i} |100\rangle  +0.159 e^{0.068 i} |101\rangle + 0.206 e^{2.671 i} |110\rangle + 0.208 e^{3.087 i} |111\rangle. \nonumber
\end{eqnarray}

\section{Three-qubit Bell inequalities}
\label{appendixA}

Below we present the the most relevant three-qubit Bell inequalities with respect to the nonlocal fraction.

\begin{table}[H]
\caption{A set of Bell inequalities for the $2 \times 2 \times 2$ scenario  \cite{Pitowskypra64_2001} which lead to the highest nonlocal fraction.}
\label{tab:tab_222}
%\begin{ruledtabular}
\begin{tabular}{l}
%\begin{eqnarray}
%\hline
%$ I_4 = 2E^{(3)}_0 - E^{(1)}_0E^{(2)}_0E^{(3)}_0 - E^{(1)}_1E^{(2)}_0E^{(3)}_0 - E^{(1)}_0E^{(2)}_1E^{(3)}_0 + E^{(1)}_1E^{(2)}_1E^{(3)}_0 + E^{(1)}_0E^{(2)}_0 + E^{(1)}_1E^{(2)}_0 + E^{(1)}_0E^{(2)}_1 - E^{(1)}_1E^{(2)}_1 - 2$%~~~~ \nonumber\\
%\vspace{2mm}\\
$I_5=E^{(1)}_0 - E^{(1)}_0E^{(2)}_0E^{(3)}_0 - E^{(1)}_0E^{(2)}_0E^{(3)}_1 + E^{(1)}_0E^{(2)}_1 - E^{(1)}_0E^{(2)}_1E^{(3)}_0 + E^{(1)}_0E^{(3)}_1 + E^{(1)}_1E^{(2)}_0 - E^{(1)}_1E^{(2)}_0E^{(3)}_0 $\\ 
$~~~~~~- E^{(1)}_1E^{(2)}_1+ E^{(1)}_1E^{(2)}_1E^{(3)}_1 + E^{(1)}_1E^{(3)}_0 - E^{(1)}_1E^{(3)}_1 + E^{(2)}_0 + E^{(2)}_0E^{(3)}_1 + E^{(2)}_1E^{(3)}_0 - E^{(2)}_1E^{(3)}_1 + E^{(3)}_0 - 3$
\vspace{2mm}\\ % ~~~~I_5\nonumber\\
$I_6=E^{(1)}_0 + E^{(1)}_0E^{(2)}_0 - E^{(1)}_0E^{(2)}_0E^{(3)}_0 - E^{(1)}_0E^{(2)}_1E^{(3)}_0 - E^{(1)}_0E^{(2)}_1E^{(3)}_1 + E^{(1)}_0E^{(3)}_1 - E^{(1)}_1E^{(2)}_0E^{(3)}_0 $\\ 
$~~~~~~ + E^{(1)}_1E^{(2)}_0E^{(3)}_1 + E^{(1)}_1E^{(3)}_0 - E^{(1)}_1E^{(3)}_1 + E^{(2)}_0 - E^{(2)}_0E^{(3)}_1 + E^{(2)}_1E^{(3)}_0 + E^{(2)}_1E^{(3)}_1 + E^{(3)}_0 - 3$
\vspace{2mm}\\
$I_{13}=2E^{(1)}_0E^{(2)}_0 + E^{(1)}_0E^{(2)}_0E^{(3)}_0 + E^{(1)}_0E^{(2)}_0E^{(3)}_1 + E^{(1)}_0E^{(2)}_1E^{(3)}_0 - E^{(1)}_0E^{(2)}_1E^{(3)}_1 + 2E^{(1)}_1E^{(2)}_0 - E^{(1)}_1E^{(2)}_0E^{(3)}_0$\\ 
$~~~~~~ - E^{(1)}_1E^{(2)}_0E^{(3)}_1 - E^{(1)}_1E^{(2)}_1E^{(3)}_0 + E^{(1)}_1E^{(2)}_1E^{(3)}_1 - 4$% ~~~~I_{_13}\nonumber\\
\vspace{2mm}\\
$I_{16}=E^{(1)}_0 + E^{(1)}_0E^{(2)}_0 + E^{(1)}_0E^{(2)}_0E^{(3)}_1 + E^{(1)}_0E^{(2)}_1E^{(3)}_0 - E^{(1)}_0E^{(2)}_1E^{(3)}_1 + E^{(1)}_0E^{(3)}_0 + E^{(1)}_1 + E^{(1)}_1E^{(2)}_0$\\ 
$~~~~~~  - 2E^{(1)}_1E^{(2)}_0E^{(3)}_0 - E^{(1)}_1E^{(2)}_0E^{(3)}_1 - E^{(1)}_1E^{(2)}_1E^{(3)}_0 + E^{(1)}_1E^{(2)}_1E^{(3)}_1 + E^{(1)}_1E^{(3)}_0 - 4$% ~~~~I_{_16}\nonumber\\
\vspace{2mm}\\
$I_{19}=E^{(1)}_0 + E^{(1)}_0E^{(2)}_0 + E^{(1)}_0E^{(2)}_0E^{(3)}_1 - E^{(1)}_0E^{(2)}_1E^{(3)}_0 + E^{(1)}_0E^{(2)}_1E^{(3)}_1 + E^{(1)}_0E^{(3)}_0 + E^{(1)}_1 + E^{(1)}_1E^{(2)}_0 $\\
$~~~~~~ - E^{(1)}_1E^{(2)}_0E^{(3)}_1 - E^{(1)}_1E^{(2)}_1E^{(3)}_0 - E^{(1)}_1E^{(2)}_1E^{(3)}_1 + E^{(1)}_1E^{(3)}_0 - 2E^{(2)}_0E^{(3)}_0 + 2E^{(2)}_1E^{(3)}_0 - 4$
\vspace{2mm}\\
$I_{21}=E^{(1)}_0 + E^{(1)}_0E^{(2)}_0 - 2E^{(1)}_0E^{(2)}_0E^{(3)}_0 + E^{(1)}_0E^{(2)}_0E^{(3)}_1 - E^{(1)}_0E^{(2)}_1E^{(3)}_0 - E^{(1)}_0E^{(2)}_1E^{(3)}_1 + E^{(1)}_0E^{(3)}_0 + E^{(1)}_1 $\\ 
$~~~~~~ + E^{(2)}_1 - E^{(1)}_1E^{(2)}_0E^{(3)}_0 - E^{(1)}_1E^{(2)}_0E^{(3)}_1 - E^{(1)}_1E^{(2)}_1 + E^{(1)}_1E^{(2)}_1E^{(3)}_1 + E^{(1)}_1E^{(3)}_0 + E^{(2)}_0 + E^{(2)}_0E^{(3)}_0 $\\
$~~~~~~  + E^{(2)}_1E^{(3)}_0 - 4$
\vspace{2mm}\\
$I_{30}=E^{(1)}_0 + 2E^{(1)}_0E^{(2)}_0 + 2E^{(1)}_0E^{(2)}_0E^{(3)}_0 - 2E^{(1)}_0E^{(2)}_0E^{(3)}_1 + E^{(1)}_0E^{(2)}_1 + E^{(1)}_0E^{(2)}_1E^{(3)}_0 + 2E^{(1)}_0E^{(2)}_1E^{(3)}_1 $\\ 
$~~~~~~ + E^{(1)}_0E^{(3)}_0 - 2E^{(1)}_1E^{(2)}_0 + E^{(1)}_1E^{(2)}_0E^{(3)}_0 - E^{(1)}_1E^{(2)}_0E^{(3)}_1 - E^{(1)}_1E^{(2)}_1 + 2E^{(1)}_1E^{(2)}_1E^{(3)}_0 + E^{(1)}_1E^{(2)}_1E^{(3)}_1 $\\ 
$~~~~~~  + E^{(1)}_1 - E^{(1)}_1E^{(3)}_0 - E^{(2)}_0E^{(3)}_0 + E^{(2)}_0E^{(3)}_1 - E^{(2)}_1E^{(3)}_0 - E^{(2)}_1E^{(3)}_1 - 6$
%\end{eqnarray}
\end{tabular}
%\end{ruledtabular}
\end{table}

%\newpage
\begin{table}[H]
\caption{A set of $m_1 \times m_2 \times m_3$ Bell inequalities, which were the most frequently violated for given (random) measurement settings, identified by mean of a linear programming method.
}
\label{tab:tab_322}
%\begin{ruledtabular}
\begin{tabular}{@{}l@{}}
$I^{322}_1=-2 E^{(1)}_0E^{(2)}_0E^{(3)}_0 -  2 E^{(1)}_0E^{(2)}_1E^{(3)}_1 + E^{(1)}_1E^{(2)}_0E^{(3)}_0 - E^{(1)}_1E^{(2)}_0E^{(3)}_1 + E^{(1)}_1E^{(2)}_1E^{(3)}_0 - E^{(1)}_1E^{(2)}_1E^{(3)}_1 $ \\
 $~~~~~~- E^{(1)}_2E^{(2)}_0E^{(3)}_0 - E^{(1)}_2E^{(2)}_0E^{(3)}_1 + E^{(1)}_2E^{(2)}_1E^{(3)}_0+ E^{(1)}_2E^{(2)}_1E^{(3)}_1- 4$
\vspace{1.5mm}\\ 
$I^{322}_2= -E^{(1)}_0 + E^{(1)}_0E^{(2)}_0E^{(3)}_0 + E^{(1)}_0E^{(2)}_1 - E^{(1)}_0E^{(2)}_1E^{(3)}_1 - E^{(1)}_0E^{(3)}_1 - E^{(1)}_1E^{(2)}_0E^{(3)}_1 - E^{(1)}_1E^{(2)}_1E^{(3)}_0 + E^{(1)}_2 $ \\
 $~~~~~~ - E^{(1)}_2E^{(2)}_0E^{(3)}_0 + E^{(1)}_2E^{(2)}_1 + E^{(1)}_2E^{(2)}_1E^{(3)}_1 - E^{(1)}_2E^{(3)}_1 + E^{(2)}_0E^{(3)}_0- 4$
\vspace{1.5mm}\\ 
$I^{322}_3= E^{(1)}_0E^{(2)}_0E^{(3)}_0 - E^{(1)}_0E^{(2)}_0E^{(3)}_1 + E^{(1)}_0E^{(2)}_1E^{(3)}_0 + E^{(1)}_0E^{(2)}_1E^{(3)}_1 + 2 E^{(1)}_0E^{(3)}_0 + E^{(1)}_1E^{(2)}_1E^{(3)}_0 + E^{(1)}_1E^{(2)}_1E^{(3)}_1 $ \\
 $~~~~~~- E^{(1)}_1E^{(3)}_0 - E^{(1)}_1E^{(3)}_1 - E^{(1)}_2E^{(2)}_0E^{(3)}_0 + E^{(1)}_2E^{(2)}_0E^{(3)}_1 + E^{(1)}_2E^{(3)}_0 - E^{(1)}_2E^{(3)}_1- 4$
\vspace{1.5mm}\\ 
$I^{322}_4= E^{(1)}_0E^{(2)}_0 + 2 E^{(1)}_0E^{(2)}_0E^{(3)}_0 + E^{(1)}_0E^{(2)}_0E^{(3)}_1 + E^{(1)}_0E^{(2)}_1 - E^{(1)}_0E^{(2)}_1E^{(3)}_1 - E^{(1)}_1E^{(2)}_0 + E^{(1)}_1E^{(2)}_0E^{(3)}_0 $ \\
 $~~~~~~- E^{(1)}_1E^{(2)}_1 + E^{(1)}_1E^{(2)}_1E^{(3)}_0 + E^{(1)}_2E^{(2)}_0E^{(3)}_0 - E^{(1)}_2E^{(2)}_0E^{(3)}_1 - E^{(1)}_2E^{(2)}_1E^{(3)}_0 + E^{(1)}_2E^{(2)}_1E^{(3)}_1- 4$
\vspace{1.5mm}\\ 
$I^{322}_5=-E^{(1)}_0E^{(2)}_0E^{(3)}_0 + E^{(1)}_0E^{(2)}_0E^{(3)}_1 + E^{(1)}_0E^{(3)}_0 - E^{(1)}_0E^{(3)}_1 + E^{(1)}_1E^{(2)}_0 - E^{(1)}_1E^{(2)}_0E^{(3)}_0 + E^{(1)}_1E^{(2)}_1+ E^{(2)}_1 $ \\
 $~~~~~~ + E^{(1)}_1E^{(2)}_1E^{(3)}_0 + 2 E^{(1)}_1E^{(3)}_1 + E^{(1)}_2E^{(2)}_1E^{(3)}_0 + E^{(1)}_2E^{(2)}_1E^{(3)}_1- E^{(1)}_2E^{(3)}_0 - E^{(1)}_2E^{(3)}_1 + E^{(2)}_0- E^{(2)}_0E^{(3)}_1$ \\
 $~~~~~~ - E^{(2)}_1E^{(3)}_1- 4$
\vspace{1.5mm}\\ 
$I^{322}_6=2 E^{(1)}_0E^{(2)}_1E^{(3)}_0 - 2 E^{(1)}_0E^{(2)}_1E^{(3)}_1 - E^{(1)}_1E^{(3)}_0 - E^{(1)}_1E^{(3)}_1 - E^{(1)}_1E^{(2)}_0E^{(3)}_0 - E^{(1)}_1E^{(2)}_0E^{(3)}_1 + E^{(1)}_2E^{(3)}_0 $ \\
 $~~~~~~ + E^{(1)}_2E^{(3)}_1 - E^{(1)}_2E^{(2)}_0E^{(3)}_0 - E^{(1)}_2E^{(2)}_0E^{(3)}_1- 4$
\vspace{1.5mm}\\ 
$I^{322}_7=-E^{(1)}_0E^{(2)}_0E^{(3)}_0 - E^{(1)}_0E^{(2)}_0E^{(3)}_1 - E^{(1)}_0E^{(2)}_1E^{(3)}_0 - E^{(1)}_0E^{(2)}_1E^{(3)}_1 - E^{(1)}_1E^{(2)}_0E^{(3)}_0 + E^{(1)}_1E^{(2)}_0E^{(3)}_1  $\\
 $~~~~~~ - E^{(1)}_1E^{(2)}_1E^{(3)}_0 + E^{(1)}_1E^{(2)}_1E^{(3)}_1- 2 E^{(1)}_2E^{(2)}_0 + 2 E^{(1)}_2E^{(2)}_1- 4$
\vspace{1.5mm}\\ 
$I^{322}_8=E^{(1)}_0 - E^{(1)}_0E^{(2)}_0  + E^{(1)}_0E^{(3)}_0 - E^{(1)}_0E^{(2)}_0E^{(3)}_1 - E^{(1)}_0E^{(2)}_1E^{(3)}_0 + E^{(1)}_0E^{(2)}_1E^{(3)}_1 + E^{(1)}_1E^{(2)}_0  - E^{(1)}_1E^{(2)}_0E^{(3)}_0$\\
 $~~~~~~+ E^{(1)}_1E^{(2)}_1 - E^{(1)}_1E^{(2)}_1E^{(3)}_0 - E^{(1)}_2 - E^{(1)}_2E^{(2)}_1- E^{(1)}_2E^{(3)}_0 - E^{(1)}_2E^{(2)}_0E^{(3)}_0 - E^{(1)}_2E^{(2)}_0E^{(3)}_1 $\\
 $~~~~~~+ E^{(1)}_2E^{(2)}_1E^{(3)}_1- 4$
\vspace{1.5mm}\\ 
$I^{322}_9=E^{(1)}_0E^{(2)}_1E^{(3)}_0 + E^{(1)}_0E^{(2)}_1E^{(3)}_1 - E^{(1)}_1E^{(2)}_1 - E^{(1)}_1E^{(2)}_1E^{(3)}_0 + E^{(1)}_2E^{(2)}_1 - E^{(1)}_2E^{(2)}_1E^{(3)}_1 - E^{(1)}_0E^{(2)}_2E^{(3)}_0$ \\
 $~~~~~~ - E^{(1)}_0E^{(2)}_2E^{(3)}_1- E^{(1)}_1E^{(2)}_2 - E^{(1)}_1E^{(2)}_2E^{(3)}_1+ E^{(1)}_2E^{(2)}_2 - E^{(1)}_2E^{(2)}_2E^{(3)}_0 - E^{(1)}_1E^{(3)}_0 + E^{(1)}_1E^{(3)}_1 $\\
  $~~~~~~ - E^{(1)}_2E^{(3)}_0 + E^{(1)}_2E^{(3)}_1- 4$
\vspace{1.5mm}\\ 
%\hline\\
$I^{332}_1=-E^{(1)}_0E^{(2)}_0 + E^{(1)}_0E^{(2)}_0E^{(3)}_0 - 2 E^{(1)}_0E^{(2)}_1E^{(3)}_1 + E^{(1)}_0E^{(2)}_2 + E^{(1)}_0E^{(2)}_2E^{(3)}_0 + 2 E^{(1)}_1E^{(2)}_0E^{(3)}_1+ E^{(1)}_2E^{(2)}_0$\\
 $~~~~~~ + 4 E^{(1)}_1E^{(2)}_1E^{(3)}_0 + 2 E^{(1)}_1E^{(2)}_2E^{(3)}_1 + E^{(1)}_2E^{(2)}_0E^{(3)}_0 - 2 E^{(1)}_2E^{(2)}_1E^{(3)}_1 - E^{(1)}_2E^{(2)}_2 + E^{(1)}_2E^{(2)}_2E^{(3)}_0- 8$ 
\vspace{1.5mm}\\ 
$I^{332}_2=-E^{(1)}_0E^{(2)}_0E^{(3)}_0 - E^{(1)}_0E^{(2)}_0E^{(3)}_1 - E^{(1)}_0E^{(2)}_1E^{(3)}_0 + E^{(1)}_0E^{(2)}_1E^{(3)}_1 + 2 E^{(1)}_0E^{(2)}_2E^{(3)}_1 - E^{(1)}_1E^{(2)}_1E^{(3)}_0$\\
 $~~~~~~  + E^{(1)}_1E^{(2)}_1E^{(3)}_1 + E^{(1)}_1E^{(2)}_2E^{(3)}_0 - E^{(1)}_1E^{(2)}_2E^{(3)}_1 - E^{(1)}_2E^{(2)}_0E^{(3)}_0 - E^{(1)}_2E^{(2)}_0E^{(3)}_1 - E^{(1)}_2E^{(2)}_2E^{(3)}_0$\\
 $~~~~~~  - E^{(1)}_2E^{(2)}_2E^{(3)}_1- 4$
\vspace{1.5mm}\\ 
$I^{332}_3=-E^{(1)}_0E^{(2)}_0 + E^{(1)}_0E^{(2)}_0E^{(3)}_0 + E^{(1)}_0E^{(2)}_1E^{(3)}_0 - E^{(1)}_0E^{(2)}_1E^{(3)}_1 - E^{(1)}_0E^{(2)}_2 + E^{(1)}_0E^{(2)}_2E^{(3)}_1 + E^{(1)}_1E^{(2)}_0$\\
 $~~~~~~  - E^{(1)}_1E^{(2)}_0E^{(3)}_0 - E^{(1)}_1E^{(2)}_1 - E^{(1)}_1E^{(2)}_1E^{(3)}_1+ E^{(1)}_1E^{(2)}_2E^{(3)}_0 + E^{(1)}_1E^{(2)}_2E^{(3)}_1 - E^{(1)}_2E^{(2)}_1 - E^{(1)}_2E^{(2)}_1E^{(3)}_0 $\\
 $~~~~~~ - E^{(1)}_2E^{(2)}_2 - E^{(1)}_2E^{(2)}_2E^{(3)}_0- 4$
\vspace{1.5mm}\\ 
$I^{332}_4=E^{(1)}_0E^{(2)}_0E^{(3)}_0 + E^{(1)}_0E^{(2)}_0E^{(3)}_1 + E^{(1)}_0E^{(2)}_1E^{(3)}_0 + E^{(1)}_0E^{(2)}_1E^{(3)}_1 - E^{(1)}_1E^{(2)}_0 + E^{(1)}_1E^{(2)}_0E^{(3)}_0 + E^{(1)}_1E^{(2)}_1 $\\
 $~~~~~~ - E^{(1)}_1E^{(2)}_1E^{(3)}_1 + E^{(1)}_1E^{(2)}_2E^{(3)}_0 - E^{(1)}_1E^{(2)}_2E^{(3)}_1+ E^{(1)}_2E^{(2)}_0 + E^{(1)}_2E^{(2)}_0E^{(3)}_1 - E^{(1)}_2E^{(2)}_1 - E^{(1)}_2E^{(2)}_1E^{(3)}_0$\\
 $~~~~~~  + E^{(1)}_2E^{(2)}_2E^{(3)}_0 - E^{(1)}_2E^{(2)}_2E^{(3)}_1- 4$
\vspace{1.5mm}\\ 
$I^{332}_{5}=2 E^{(1)}_0E^{(2)}_2E^{(3)}_0 - E^{(1)}_0E^{(2)}_0E^{(3)}_1 + E^{(1)}_0E^{(2)}_2E^{(3)}_1 - E^{(1)}_0E^{(2)}_0 + 2 E^{(1)}_0E^{(2)}_1 + E^{(1)}_0E^{(2)}_2 + 2 E^{(1)}_1E^{(2)}_0E^{(3)}_0 $\\
 $~~~~~~ + 2 E^{(1)}_1E^{(2)}_1E^{(3)}_0 - 2 E^{(1)}_1E^{(2)}_1E^{(3)}_1+ 2 E^{(1)}_1E^{(2)}_2E^{(3)}_1 + 2 E^{(1)}_2E^{(2)}_2E^{(3)}_0 - E^{(1)}_2E^{(2)}_0E^{(3)}_1 + E^{(1)}_2E^{(2)}_2E^{(3)}_1$\\
 $~~~~~~ + E^{(1)}_2E^{(2)}_0 - 2 E^{(1)}_2E^{(2)}_1 - E^{(1)}_2E^{(2)}_2- 8$
\vspace{1.5mm}\\ 
$I^{332}_{6}=E^{(1)}_0E^{(2)}_0E^{(3)}_0 - E^{(1)}_0E^{(2)}_0E^{(3)}_2 - E^{(1)}_0E^{(2)}_1E^{(3)}_0 + E^{(1)}_0E^{(2)}_1E^{(3)}_2 + E^{(1)}_1E^{(2)}_0E^{(3)}_0 - E^{(1)}_1E^{(2)}_0E^{(3)}_2$\\
 $~~~~~~ - E^{(1)}_1E^{(2)}_1E^{(3)}_2 + E^{(1)}_2E^{(2)}_2E^{(3)}_0+ E^{(1)}_2E^{(2)}_2E^{(3)}_2 - E^{(1)}_2E^{(3)}_0 - E^{(1)}_2E^{(3)}_2 - E^{(2)}_2E^{(3)}_0 - E^{(2)}_2E^{(3)}_2$\\
 $~~~~~~  + E^{(1)}_1E^{(2)}_1E^{(3)}_0  - E^{(3)}_0 - E^{(3)}_2- 4$
\vspace{1.5mm} \\
%\hline\\
$I^{333}_{1}=E^{(1)}_0E^{(2)}_0E^{(3)}_0 + E^{(1)}_0E^{(2)}_0E^{(3)}_1 + E^{(1)}_0E^{(2)}_2E^{(3)}_0 + E^{(1)}_0E^{(2)}_2E^{(3)}_1 - E^{(1)}_1E^{(2)}_0E^{(3)}_0 - E^{(1)}_1E^{(2)}_0E^{(3)}_2 $\\
 $~~~~~~ - E^{(1)}_1E^{(2)}_1E^{(3)}_1 + E^{(1)}_1E^{(2)}_2E^{(3)}_1+ E^{(1)}_1E^{(2)}_2E^{(3)}_2 - E^{(1)}_2E^{(2)}_0E^{(3)}_1 + E^{(1)}_2E^{(2)}_0E^{(3)}_2 + E^{(1)}_2E^{(2)}_1E^{(3)}_0$\\
 $~~~~~~ + E^{(1)}_1E^{(2)}_1E^{(3)}_0 - E^{(1)}_2E^{(2)}_1E^{(3)}_1 + E^{(1)}_2E^{(2)}_2E^{(3)}_0 - E^{(1)}_2E^{(2)}_2E^{(3)}_2- 4$
\vspace{1.5mm}\\ 
$I^{333}_{2}=E^{(1)}_0E^{(2)}_0E^{(3)}_0 - E^{(1)}_0E^{(2)}_0E^{(3)}_2 + E^{(1)}_0E^{(2)}_1E^{(3)}_1 - E^{(1)}_0E^{(2)}_1E^{(3)}_2 + E^{(1)}_0E^{(2)}_2E^{(3)}_0 - E^{(1)}_0E^{(2)}_2E^{(3)}_1$\\
 $~~~~~~ + E^{(1)}_1E^{(2)}_0E^{(3)}_2 + 2 E^{(1)}_1E^{(2)}_1E^{(3)}_0+ E^{(1)}_1E^{(2)}_1E^{(3)}_1 + E^{(1)}_1E^{(2)}_1E^{(3)}_2 + E^{(1)}_1E^{(2)}_2E^{(3)}_0 - E^{(1)}_1E^{(2)}_0E^{(3)}_0$\\
 $~~~~~~ - E^{(1)}_1E^{(2)}_2E^{(3)}_1- 4$
\end{tabular}
%\end{ruledtabular}
\end{table}

\end{document}